\documentclass[journal=jpcbfk,manuscript=article,layout=traditional]{achemso}

\usepackage{graphicx}
\usepackage{dcolumn}
\usepackage{bm}
\usepackage{booktabs}
\usepackage{mathtools}
\usepackage{color}
\usepackage{amssymb}

\usepackage[version=3]{mhchem} 
\usepackage[T1]{fontenc}       

\author{Anthony R. Poggioli}
\affiliation
{Laboratoire de Physique Statistique, Ecole Normale Sup\'{e}rieure, Paris}
\alsoaffiliation
{Centre National de la Recherche Scientifique}
\author{Alessandro Siria}
\affiliation
{Laboratoire de Physique Statistique, Ecole Normale Sup\'{e}rieure, Paris}
\alsoaffiliation
{Centre National de la Recherche Scientifique}
\author{Lyd\'{e}ric Bocquet}
\affiliation
{Laboratoire de Physique Statistique, Ecole Normale Sup\'{e}rieure, Paris}
\alsoaffiliation
{Centre National de la Recherche Scientifique}
\email{lyderic.bocquet@lps.ens.fr}

\title[An \textsf{achemso} demo]
  {Beyond the Trade-Off: Dynamic Selectivity in Ionic Transport and Current Rectification}

\begin{document}

\begin{tocentry}

\includegraphics{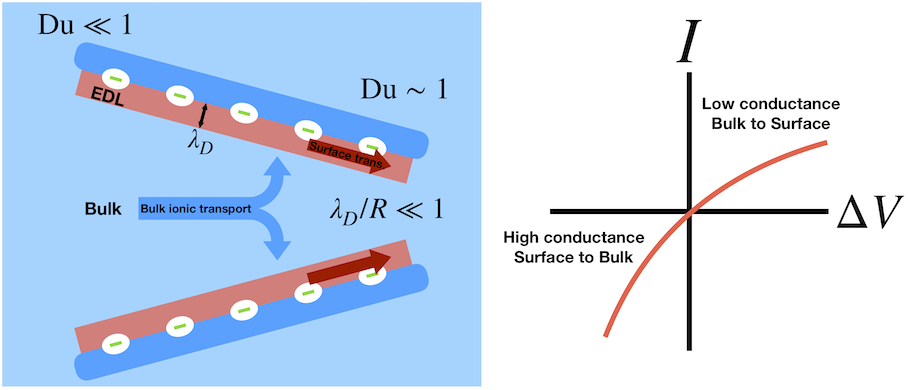}

\end{tocentry}

\begin{abstract}
Traditionally, ion-selectivity in nanopores and nanoporous membranes is understood to be a consequence of Debye overlap, in which the Debye screening length is comparable to the nanopore radius somewhere along the length of the nanopore(s).  This criterion sets a significant limitation on the size of ion-selective nanopores, as the Debye length is on the order of $1 - 10$ nm for typical ionic concentrations.  However, the analytical results we present here demonstrate that surface conductance generates a dynamical selectivity in  ion transport, and this selectivity is controlled by so-called Dukhin, rather than Debye, overlap.  The Dukhin length, defined as the ratio of surface to bulk conductance, reaches values of hundreds of nanometers for typical surface charge densities and ionic concentrations, suggesting the possibility of designing large-nanopore ($10 - 100$ nm), high-conductance membranes exhibiting significant ion-selectivity.  Such membranes would have potentially dramatic implications for the efficiency of osmotic energy conversion and separation techniques.  Furthermore, we demonstrate that this mechanism of dynamic selectivity leads ultimately to the rectification of ionic current, rationalizing previous studies showing that Debye overlap is not a necessary condition for the occurrence of rectifying behavior in nanopores.
\end{abstract}

\section{Introduction}

Ionic and water transport in nanometric confinement has been an active topic of research for two decades \citep[][]{Bocquet+Charlaix2010}, with practical applications to macro-/bio-molecular analysis \citep{Misakian+Kasianowicz2003, Schiedt_et_al2005, Vlassiouk_et_al2009, Zhou_et_al2011}, energy generation \citep{Bocquet+Charlaix2010,Siria_et_al2017,Rankin+Huang2016,Ren+Stein2008,Siria_et_al2013,van_der_Heyden_et_al2006}, and desalination \citep{Bocquet+Charlaix2010, Picallo_et_al2013}.  Of particular interest is understanding via what mechanisms and under what conditions one may obtain nonlinear responses of fluxes (charge, solute mass, solvent mass) to external forcings (voltage, pressure, concentration difference).  A detailed theoretical understanding of such nonlinear transport processes is a necessary first step in designing exotic tailored nanopore and membrane functionalities \citep{Jubin_et_al2018}.  The most well-known example of such a nonlinear response is ionic current rectification (ICR), in which the ionic currents driven through nanopores by applied voltages of equal magnitude and opposite sign are found to be of unequal magnitude, in analogy toff solid-state semiconductor diodes \citep{Shockley1949, Bocquet+Charlaix2010}.  ICR has been extensively studied experimentally and via continuum simulations, \citep[e.g.][]{Ai_et_al2010, Cervera_et_al2006, Constantin+Siwy2007, Liu_et_al2007, Nguyen_et_al2010, Siwy+Fulinski2002, Wang_et_al2007, Kubeil+Bund2011, Lan_et_al2011, Laohakunakorn+Keyser2015, White+Bund2008, Kovarik_et_al2009, Li_et_al2013}, though comparatively few studies have examined the phenomenon analytically \citep{Picallo_et_al2013,Siwy+Fulinski2002,Vlassiouk_et_al_acsnano2008}.  Experimentally, it is found that ICR may be induced by unequal reservoir concentrations \citep{Cheng+Guo2007}, asymmetric geometries \citep{Cervera_et_al2006, Perry_et_al2010, Siwy+Fulinski2002, Vlassiouk+Siwy2007, Kovarik_et_al2009, White_et_al2006}, or asymmetric surface charge distributions \citep{Karnik_et_al2007, Nguyen_et_al2010, Li_et_al2013}.

ICR is generally understood to be a consequence of the accumulation or depletion of ionic concentration induced by a gradient in ion selectivity along the length of a nanopore \citep{Woermann2003}.  The key parameter controlling the local selectivity is accordingly the ratio of the local Debye length $\lambda_D$ to the local nanopore radius $R$.  The Debye length is defined as
\begin{equation}
\lambda_D \equiv \sqrt{\frac{\epsilon_r \epsilon_0 k_B T}{e^2 \sum_j z_j^2 c_j}},
\label{eqn:Debye_length}
\end{equation}
where $\epsilon_r$ and $\epsilon_0$ are, respectively, the relative dielectric permittivity of the solvent and the vacuum permittivity, $k_B$ is the Boltzmann constant, $T$ is the thermodynamic temperature, $e$ is the elementary charge, $z_j$ is the valence of the $j^{\rm th}$ ionic species, and $c_j$ is the concentration of the $j^{\rm th}$ ionic species.  This length scale characterizes the thickness of the diffuse layer of net ionic charge that forms in the vicinity of charged surfaces \citep{Bocquet+Charlaix2010}.  As the diffuse layer must counterbalance the surface charge, a region of strong Debye layer overlap, $\lambda_D/R \gtrsim \mathcal{O}(1)$, should be highly selective to counterions, while a region of weak Debye overlap, $\lambda_D/R \ll 1$, should be essentially non-selective.

Debye overlap has significant consequences for ionic transport through the pore. In order to illustrate the underlying physical mechanisms, consider a conical nanopore with a uniform surface charge density, as discussed in \citep{Woermann2003}. When $\lambda_D/R \gtrsim \mathcal{O}(1)$ in the vicinity of the nanopore tip, a Donnan equilibrium between the tip and the connected reservoir forms, and this region is highly selective to (positive) counterions.  The tip is then a region of increased transference of cations and suppressed transference of anions.  Thus, when an external electric field is directed from the (non-selective) base to the (highly selective) tip, both cations and anions pass from regions of relatively lower to relatively higher transference, and the result at steady state is a depletion of ionic concentration in the nanopore interior.  The depletion of charge carriers results in a decrease of the local electrophoretic conductivity and corresponds to the low conductance (reverse-bias) diode state.  On the other hand, if the direction of the applied field is inverted, both ionic species pass from regions of relatively high to relatively low transference, resulting in an accumulation of ionic concentration.  This accumulation of charge carriers likewise results in a high conductance (forward-bias) state. Altogether, the criterion $\lambda_D/R \gtrsim 1$ is expected to be a prerequisite to observe non-linear ionic transport, and in particular ionic current rectification. The mechanism of concentration accumulation and depletion in ICR has since been extensively corroborated numerically \citep{Ai_et_al2010, Vlassiouk_et_al_acsnano2008, Cervera_et_al2006, Constantin+Siwy2007, Liu_et_al2007, Nguyen_et_al2010, Cheng+Guo2010, Kubeil+Bund2011, Lan_et_al2011, Laohakunakorn+Keyser2015, White+Bund2008} and experimentally \citep{Perry_et_al2010}.  

However, recent studies have indicated not only that the parameter $\lambda_D/R$ fails to predict the occurrence or strength of rectification, but also that significant rectification may be obtained even when the Debye length is one-to-three orders of magnitude smaller than the minimum nanopore radius \citep{Kovarik_et_al2009, Kubeil+Bund2011, Lan_et_al2011, Laohakunakorn+Keyser2015, White+Bund2008, White_et_al2006, Jubin_et_al2018, He_et_al2017, Lin_et_al2018}.  To our knowledge, no consistent alternative criterion has been proposed to predict the occurrence of ICR.  Furthermore, while there have been many numerical simulations of ICR within the Poisson-Nernst-Planck (PNP) or Poisson-Nernst-Planck-Stokes (PNPS) framework, few studies have offered a detailed theoretical analysis of the phenomenon, and these studies are typically confined to quite specialized scenarios, \citep[e.g.][]{Picallo_et_al2013}.

In this paper, we demonstrate that the nanopore selectivity is not determined by the relative value of the Debye length $\lambda_D$ compared to the pore radius, but rather by a dynamic criterion related to the relative magnitudes of the surface and bulk ionic conductions. This introduces the so-called Dukhin length $\ell_{\rm Du}$, defined as the ratio of the surface conductivity to the bulk conductivity in the nanopore \citep{Bocquet+Charlaix2010}. The Dukhin length can be adequatly rewritten in terms of the charge density on the pore surface, $\sigma$, as $\ell_{\rm Du}= (|\sigma|/e)/c$, with $c$ the bulk salt concentration \citep{Bocquet+Charlaix2010}. A dimensionless Dukhin number is accordingly introduced as
\begin{equation}
{\rm Du} \equiv \frac{\ell_{\rm Du}}{R} \equiv \frac{|\sigma|}{e c R}.
\label{eqn:Dukhin}
\end{equation}
The Dukhin length approaches values of hundreds of nanometers for typical surface charges in the range of $10-100$ mC m$^{-2}$ and concentrations in the range of $0.1 - 1$ mM, one-to-two orders of magnitude larger than the corresponding Debye lengths. Written in terms of the surface charge density (Eq. \ref{eqn:Dukhin}), we note that the criterion Du $> 1$ for substantial selectivity/ICR is equivalent to the criterion $|\sigma| > e c R$ noted by some authors \citep{Vlassiouk_et_al2008_nanolett, Stein_et_al2004, Vlassiouk+Siwy2007}.

As we demonstrate below, substantial ionic selectivity may be obtained when the nanopore radius is comparable to the Dukhin length.  This is consistent with numerical results indicating substantial ion-selectivity may be obtained in highly charged pores with radii much larger than the Debye length \citep{Vlassiouk_et_al2008_nanolett}, and it is in stark contrast to traditional ion-selective membranes, which typically have subnanometric pore sizes \citep{Siria_et_al2017}.  We term this mechanism dynamic selectivity, in contrast to the 'thermodynamic' picture of ionic selectivity based on Debye overlap and the formation of a Donnan equilibrium at the ends of the nanopore, \citep[e.g.][]{Woermann2003, Plecis_et_al2005, Bocquet+Charlaix2010}.  The possibility of obtaining significant selectivity for large ($10 - 100$ nm) pores may have significant implications for, e.g., osmotic energy generation; we will discuss this point in more detail below.

Furthermore, we will show below that a gradient in the local Dukhin number along the length of the nanopore results in a repartitioning of the fraction of the ionic transport carried in the  non-selective bulk and in the highly selective Debye layer.  It is this repartitioning, rather than Debye overlap and the formation of a local Donnan equilibrium at one end of the nanopore, that results in ICR.

Our starting point will be the derivation of one-dimensional transport equations from a radial integration of the axisymmetric PNP equations in the limit that $\lambda_D/R \ll 1$ but Du $\sim 1$; we focus on this limit as we are particularly interested in rationalizing those experimental \citep{Lan_et_al2011, Jubin_et_al2018, Kovarik_et_al2009, White_et_al2006, He_et_al2017, Lin_et_al2018} and numerical \citep{Kubeil+Bund2011, Laohakunakorn+Keyser2015, White+Bund2008, Jubin_et_al2018, Vlassiouk_et_al2008_nanolett} results indicating that substantial rectification and selectivity may occur even when $\lambda_D/R \ll 1$.  We note that a reanalysis of the data reported in the literature corroborates the assertion that significant rectification occurs when Du $\sim 1$, irrespective of the value of $\lambda_D/R$; this is discussed below.

From these transport equations, we are able to derive a simple implicit expression for the current-voltage (IV) response in a concentration diode.  To our knowledge, this is the first time such a solution has been presented.  Additionally, we will give numerical solutions of these transport equations for a geometric diode (rectification induced by an asymmetric, continuously varying radius) and a charge diode (rectification induced by an asymmetric, continuously varying surface charge density distribution).  These solutions will directly illustrate the key role of the local Dukhin number and the mechanism of dynamic selectivity in ICR.  Finally, we derive analytical expressions for general limiting conductances directly from the transport equations that will be useful in estimating, for example, surface charge densities from rectified IV curves.

\section{Theory}

\subsection{1D Transport Equations in the Absence of Debye Overlap}

In this section, we derive one-dimensional transport equations for the electrostatic potential and total ionic concentration at the nanopore centerline from the axisymmetric Poisson-Nernst-Planck (PNP) equations.  Our derivation relies on the geometric constraint $R_0/\ell \ll 1$, where $R_0$ is a scale of the radial extent of the nanopore, and $\ell$ is a characteristic scale of variation of the nanopore geometry.  Such a slowly varying approximation implies that a Poisson-Boltzmann (PB) equilibrium holds locally on each cross-section \citep{Fair+Osterle1971, Rankin+Huang2016}. As we intend to demonstrate that the Dukhin number is the principal parameter controlling selectivity and ICR, and that ICR may occur even in the absence of Debye overlap anywhere along the length of the nanopore, we focus on the regime that $\lambda_D/R \ll 1$.  There is no Debye overlap in the center of the nanopore so that the electrolyte there is electroneutral, and we may partition the ionic concentrations and electrostatic potential as follows \citep{Fair+Osterle1971}:
\begin{equation}
c_{\pm}(x,r) = \frac{c_0(x)}{2} + \delta c_{\pm}(x,r), \quad \text{and}
\label{eqn:cpm_partition}
\end{equation}
\begin{equation}
\phi(x,r) = \phi_0(x) + \delta \phi(x,r),
\label{eqn:phi_partition}
\end{equation}
where $c_0(x) \equiv c(x,0)$ is the value of the total ionic concentration $c \equiv c_+ + c_-$ at the nanopore centerline ($r = 0$), $\phi_0(x) \equiv \phi(x,0)$ is the electrostatic potential at the nanopore centerline, and $\delta c_{\pm}(x,r)$ and $\delta \phi(x,r)$ are the radial deviations in the ionic concentrations and electrostatic potential induced by the formation of a screening Debye layer in the vicinity of the nanopore wall.  In Eq. \ref{eqn:cpm_partition}, we have assumed a symmetric (z:z) salt.  In what follows, we will assume a monovalent (1:1) salt in which the cation and anion have identical mobility and diffusion coefficients.

With this notation and these assumptions, the steady-state PNP equations reduce to
\begin{equation}
j_{\pm} = -D \left( \frac{\partial c_{\pm}}{\partial x} \pm \frac{e}{k_B T} c_{\pm} \frac{\partial \phi}{\partial x} \right),
\label{eqn:PNP1_dim}
\end{equation}
 \begin{equation}
\frac{d}{dx} \left( 2 \pi \int_0^R \text{dr} \, r j_{\pm} \right) \equiv \frac{dJ_{\pm}}{dx} = 0,
\label{eqn:PNP2_dim}
\end{equation}
\begin{equation}
\epsilon_r \epsilon_0 \frac{1}{r} \frac{\partial}{\partial r} \left( r \frac{\partial \delta \phi}{\partial r} \right) + n_c \approx \epsilon_r \epsilon_0 \frac{\partial^2 \delta \phi}{\partial Z^2} + n_c = 0,
\label{eqn:PNP3_dim}
\end{equation}
\begin{equation}
c = c_0 \text{cosh} \left( \frac{e \delta \phi}{k_B T} \right), \quad \text{and}
\label{eqn:PNP4_dim}
\end{equation}
\begin{equation}
n_c = - e c_0 \text{sinh} \left( \frac{e \delta \phi}{k_B T} \right),
\label{eqn:PNP5_dim}
\end{equation}
where $D$ is the diffusion coefficient, $e$ is the elementary charge, $k_B$ is the Boltzmann constant, $T$ is the thermodynamic temperature, $R$ is the local nanopore radius, $\epsilon_r$ ($\approx 80$ for water at room temperature) is the relative permittivity of the solvent and $\epsilon_0$ is the vacuum permittivity, and $n_c \equiv e(c_+ - c_-)$ is the ionic charge density.  In Eq. \ref{eqn:PNP3_dim} we have neglected the portion of the radial gradient induced by the curvature of the nanopore wall as it is suppressed by a factor $\lambda_D/R \ll 1$ relative to $\partial_r^2$, and we have introduced the coordinate $Z \equiv R - r$.  Eq. \ref{eqn:PNP1_dim} is the ionic flux density in the along-flow ($x$) direction; Eq. \ref{eqn:PNP2_dim} is the cross-sectionally integrated continuity equation at steady state; Eq. \ref{eqn:PNP3_dim} is the Poisson equation, retaining only the radial component of the electric field divergence in accordance with the slowly varying approximation; and Eqs. \ref{eqn:PNP4_dim} and \ref{eqn:PNP5_dim} are the distributions of the total ionic concentration and ionic charge density obtained from the Boltzmann distribution, applied on the assumption of a slowly varying geometry. Finally, we have neglected advection as it is not expected to affect our conclusions \citep{Ai_et_al2010, Vlassiouk_et_al2008_nanolett}, and such an approach allows for tractable analytical derivations.

Before continuing, we introduce dimensionless rescaled variables, as listed in Table \ref{table:variables}; we have rescaled the $x$-coordinate by the total nanopore length $L$ and the surface charge $\sigma$ by a reference magnitude $|\sigma_{\text{ref}}|$, and we have introduced $\overline{c}$, the average of the reservoir concentrations, as a scale of ionic concentration in the nanopore.  In what follows, we will take the reference surface charge magnitude $|\sigma_{\rm ref}|$ to be either the magnitude of the surface charge density when it is uniform, or the maximum surface charge magnitude when the surface charge density is nonuniform.  We further recast Eqs. \ref{eqn:PNP1_dim} and \ref{eqn:PNP2_dim} in terms of the rescaled solute flux $J \equiv J_+ + J_- \equiv \int \text{dA} j$ and ionic current $I \equiv J_+ - J_- \equiv \int \text{dA} i$.  With these modifications, the governing equations become
\begin{equation}
j = -\left( \frac{\partial c}{\partial x} + n_c \frac{\partial \phi}{\partial x} \right),
\label{eqn:PNP1_nd}
\end{equation}
\begin{equation}
i = -\left( \frac{\partial n_c}{\partial x} + c \frac{\partial \phi}{\partial x} \right),
\label{eqn:PNP2_nd}
\end{equation}
\begin{equation}
\frac{dJ}{dx} = \frac{dI}{dx} = 0,
\label{eqn:PNP3_nd}
\end{equation}
\begin{equation}
\left( \frac{\lambda_D^{\text{ref}}}{R_0} \right)^2 \frac{\partial^2 \delta \phi}{\partial Z^2} + n_c = 0,
\label{eqn:PNP4_nd}
\end{equation}
\begin{equation}
c = c_0 \text{cosh} (\delta \phi), \quad \text{and}
\label{eqn:PNP5_nd}
\end{equation}
\begin{equation}
n_c = -c_0 \text{sinh}(\delta \phi),
\label{eqn:PNP6_nd}
\end{equation}
where we have introduced a reference Debye length $\lambda_D^{\text{ref}} \equiv \sqrt{k_B T \epsilon_r \epsilon_0 / e^2 \overline{c}}$, defined in terms of the mean reservoir concentration $\overline{c}$.

\begin{table}
 \begin{center}
  \begin{tabular}{ccc}
	quantity			& variable		& rescaled \\ \hline
	position			& $x$			& $x \rightarrow L x$ \\
	radius				& $R$			& $R \rightarrow R_{\rm min} R$ \\
	concentration		& $c$			& $c \rightarrow \overline{c} c$ \\
	ionic charge density		& $n_c$		& $n_c \rightarrow e \overline{c} n_c$ \\
	electrostatic potential	& $\phi$		& $\phi \rightarrow (k_B T/e) \phi$ \\
	electrochem. potential	& $\mu_{\pm}$	& $\mu_{\pm} \rightarrow k_B T \mu_{\pm}$ \\
	flux density			& $j_{\pm}$		& $j_{\pm} \rightarrow (D \overline{c}/L) j_{\pm}$ \\
	surface charge		& $\sigma$		& $\sigma \rightarrow |\sigma_{\text{ref}}| \sigma$ \\
	conductance			& $G$			& $G \rightarrow (R_{\rm min}^2/L) (e^2 D / k_B T) \overline{c}$
  \end{tabular}
  \caption{Independent and dependent variables and their rescaled dimensionless counterparts.}
  \label{table:variables}
 \end{center}
\end{table}

We differentiate Eq. \ref{eqn:PNP5_nd} (\ref{eqn:PNP6_nd}) with respect to $x$, insert the result into Eq. \ref{eqn:PNP1_nd} (\ref{eqn:PNP2_nd}), and integrate on the cross-section to obtain
\begin{equation}
\frac{J}{\pi R^2} = -\frac{dc_0}{dx} - \frac{\langle \delta c \rangle}{c_0} \frac{d c_0}{dx} - \frac{\langle n_c \rangle}{c_0} c_0 \frac{d\phi_0}{dx}, \quad \text{and}
\label{eqn:Jsol_v1}
\end{equation}
\begin{equation}
\frac{I}{\pi R^2} = -c_0 \frac{d\phi_0}{dx} - \frac{ \langle n_c \rangle }{c_0} \frac{d c_0}{dx} - \frac{\langle \delta c \rangle}{c_0} c_0 \frac{d\phi_0}{dx},
\label{eqn:I_v1}
\end{equation}
where $\langle \rangle \equiv A^{-1} \int \text{dA}$ denotes a cross-sectional average.  The integral of the charge density is set by the condition of local electroneutrality, a necessary consequence of a local PB equilibrium; the integral of $\delta c$ may be evaluated using PB equilibrium theory (Eqs. \ref{eqn:PNP4_nd}, \ref{eqn:PNP5_nd}, and \ref{eqn:PNP6_nd}) in the limit $\lambda_D/R \ll 1$.  The condition of local electroneutrality requires that
\begin{equation}
\frac{ \langle n_c \rangle }{c_0} = - 2 \text{Du}_{\text{ref}} \frac{\sigma}{R c_0} \equiv -2 \text{Du}(x),
\label{eqn:electroneutrality}
\end{equation}
where $\text{Du}_{\text{ref}} \equiv |\sigma_{\text{ref}}|/e \overline{c} R_0$ is a reference Dukhin number, and $\text{Du}(x)$ is the local Dukhin number.  We note that $\text{Du}_{\text{ref}}$ is defined to be always positive, but $\text{Du}(x)$ carries the sign of the local surface charge density.  The integral of $\delta c$ can be evaluated using Eqs. \ref{eqn:PNP4_nd} through \ref{eqn:PNP6_nd}.  The result is
\begin{equation}
\frac{ \langle \delta c \rangle }{c_0} = 4 \frac{\lambda_D}{R_0 R} \left\{ \sqrt{ \left[ \frac{\text{Du}}{2 (\lambda_D/R_0 R)} \right]^2 + 1} - 1 \right\},
\label{eqn:deltac_integral}
\end{equation}
where $\text{Du}$ is the local value of the Dukhin number, and $\lambda_D(x) \equiv \lambda_D^{\text{ref}}/\sqrt{c_0(x)}$ is the local Debye length.  In the limit $(\lambda_D/R_0R)/{\rm Du} \rightarrow 0$, this reduces to
\begin{equation}
\frac{ \langle \delta c \rangle }{c_0} = 2 |\text{Du}(x)|.
\label{eqn:deltac_integral_limit}
\end{equation}
We will consider this limit in developing an analytical solution for the concentration diode below, as we are most interested in the scenario that there is no Debye overlap ($\lambda_D(x)/R(x)R_0 \ll 1$ everywhere) but the Dukhin number is of order one (Du(x) $\sim 1$ somewhere).  In this case, the transport equations, Eqs. \ref{eqn:Jsol_v1} and \ref{eqn:I_v1}, become
\begin{equation}
\frac{J}{\pi R^2} = -\left[ \frac{dc_0}{dx} + 2 |\text{Du}(x)| \frac{dc_0}{dx} - 2\text{Du}(x) c_0 \frac{d\phi_0}{dx} \right], \quad \text{and}
\label{eqn:Jsol_final}
\end{equation}
\begin{equation}
\frac{I}{\pi R^2} = -\left[ c_0 \frac{d\phi_0}{dx} - 2\text{Du}(x) \frac{dc_0}{dx} + 2|\text{Du}(x)| c_0 \frac{d\phi_0}{dx} \right],
\label{eqn:I_final}
\end{equation}
where $I$ and $J$ are of course constant along the length of the nanopore (Eq. \ref{eqn:PNP3_nd}).  It is useful to distinguish between those terms in Eqs. \ref{eqn:Jsol_final} and \ref{eqn:I_final} that arise from transport outside of the Debye layer (bulk transport) and those arising from transport within the Debye layer (surface transport).  In the equation for the solute flux (Eq. \ref{eqn:Jsol_final}), the terms represent, from left to right, bulk diffusion, surface diffusion, and surface electrophoretic mass transport.  In the equation for the ionic current (Eq. \ref{eqn:I_final}), the terms represent bulk electrophoretic current, surface charge diffusion, and surface electrophoretic current.  We see that the local Dukhin number, which sets the cross-sectionally averaged ionic charge (Eq. \ref{eqn:electroneutrality}), as well as the cross-sectionally averaged excess concentration when $\lambda_D/R_0R \ll \text{Du}$ (Eq. \ref{eqn:deltac_integral_limit}), determines the ratio of surface to bulk transport.  Motivated by this observation, we quantify the ratio of the surface transport to the total transport by the following surface transport ratio (STR):
\begin{equation}
\text{STR}(x) \equiv \frac{2|\text{Du}(x)|}{1 + 2|\text{Du}(x)|}.
\label{eqn:STR}
\end{equation}
We see immediately that the partitioning of the transport into surface and bulk components adjusts along the length of the nanopore and is controlled locally by the Dukhin number.  In the case that there is a large asymmetry in Dukhin number on either end of the nanopore, this can result in a substantial repartitioning of the transport in the nanopore interior, the consequences of which will be explored in the following sections.

An illustrative form of Eqs. \ref{eqn:Jsol_final} and \ref{eqn:I_final} is obtained by introducing the definition of the local Dukhin number, Eq. \ref{eqn:electroneutrality}, and defining the coion and counterion fluxes $J_{\rm co/count} \equiv (J \pm {\rm S} I)/2$ and electrochemical potentials $\mu_{\rm co/count} \equiv {\rm ln} (c_0/2) \pm {\rm S} \phi$.  In the preceding definition, ${\rm S} \equiv {\rm sign} (\sigma)$ is the sign of the surface charge.  Inserting these definitions into Eqs. \ref{eqn:Jsol_final} and \ref{eqn:I_final}, we obtain
\begin{equation}
\frac{J_{\rm co}}{\pi R^2} = -\frac{c_0}{2} \frac{d \mu_{\rm co}}{dx}, \quad \rm and
\label{eqn:Jco}
\end{equation}
\begin{equation}
\frac{J_{\rm count}}{\pi R^2} = -\left( \frac{c_0}{2} + 2 {\rm Du}_{\rm ref} \frac{|\sigma|}{R} \right) \frac{d\mu_{\rm count}}{dx}.
\label{eqn:Jcount}
\end{equation}
Note that we have implicitly assumed in deriving these equations from Eqs. \ref{eqn:Jsol_final} and \ref{eqn:I_final} that the sign of the surface charge S does not change along the length of the nanopore.  If the sign of the surface charge does change at one (or several) points along the length of the nanopore, Eqs. \ref{eqn:Jco} and \ref{eqn:Jcount} may be applied in each region delineated by a discontinuity in S$(x)$, matching $c_0$ and $\phi_0$ at each discontinuity.  (More about the proper boundary conditions for Eqs. \ref{eqn:Jsol_final}/\ref{eqn:Jco} and \ref{eqn:I_final}/\ref{eqn:Jcount} will be said below.)

In Eqs. \ref{eqn:Jco} and \ref{eqn:Jcount}, we recognize $c_0/2$ as the concentration of both coions and counterions at the nanopore centerline.  Further, in Eq. \ref{eqn:Jcount}, we recognize 2Du$_{\rm ref} |\sigma|/R$ as the additional concentration accumulated in the Debye layer (Eq. \ref{eqn:deltac_integral_limit}).  The term 2Du$_{\rm ref} |\sigma|/R \times d\mu_{\rm count}/dx$ in Eq. \ref{eqn:Jcount} represents the entirety of the surface transport in the nanopore; this indicates that coions are perfectly excluded from the Debye layer in the limit $\lambda_D/R \ll {\rm Du}$ ($R$ dimensioned).  This will be shown to be the case from PB equilibrium theory and the condition of local equilibrium below.

We will employ Eqs. \ref{eqn:Jsol_final} and \ref{eqn:I_final} below to develop an implicit analytical solution for the current-voltage (IV) relationship in a concentration diode.

\subsection{Dynamic Selectivity}
\label{sec:selectivity}

\begin{figure}[h!]
    \centerline{\includegraphics[scale = 0.38]{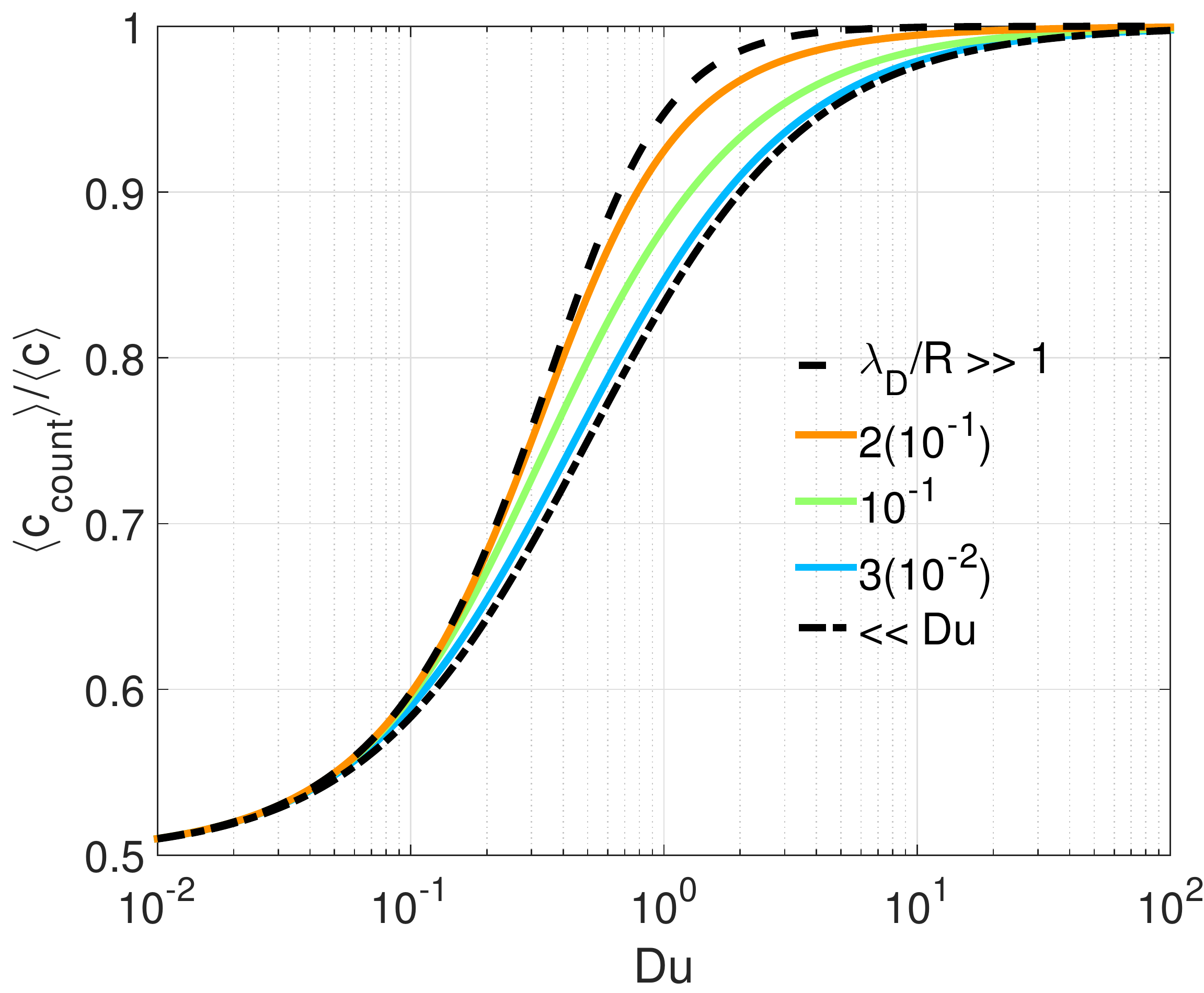}}
    \caption{The ratio of the cross-sectionally averaged counterion concentration to the cross-sectionally averaged total concentration, a metric of the nanopore selectivity, as a function of Dukhin number.  The lines are colored according to $\lambda_D/R$, as indicated in the legend.  The dashed (dot-dashed) line indicates the curve obtained in the limit $\lambda_D/R \rightarrow \infty$ ($\rightarrow 0$).}
    \label{fig:selectivity}
\end{figure}

As noted above, in the slowly varying limit ($R_0/\ell \ll 1$) considered here, the ionic concentration profiles on the cross-section deviate negligibly from those predicted by PB equilibrium theory.  Thus, we may apply this equilibrium theory to determine how the ionic selectivity on the cross-section is influenced by Debye overlap ($\lambda_D/R$) and the Dukhin number.  Using Eqs. \ref{eqn:electroneutrality} and \ref{eqn:deltac_integral}, we determine the cross-sectionally averaged total concentration, $\langle c \rangle = \langle \delta c \rangle + c_0$, and the cross-sectionally averaged counterion concentration, $\langle c_{\text{count}} \rangle = \left( \langle c \rangle + |\langle n_c \rangle| \right)/2$.  We take the ratio $\langle c_{\text{count}} \rangle/\langle c \rangle$ as a metric of selectivity; this parameter ranges between $1/2$, indicating the total average concentration is equally partitioned between coions and counterions and the nanopore is locally non-selective, and unity, indicating that the entirety of the average concentration is due to counterions and the nanopore is perfectly selective.  We find that, in the range $0 < \lambda_D/R < 2(10^{-1})$, the nanopore selectivity is strongly influenced by the local Dukhin number in the range $10^{-1} < \text{Du} < 10^1$ (Fig. \ref{fig:selectivity}).  In comparison, the influence of $\lambda_D/R$ on selectivity for this range of $\lambda_D/R$ is substantially less pronounced.

\subsubsection{From equilibrium to dynamical selectivity}
As noted above, Eqs. \ref{eqn:Jco} and \ref{eqn:Jcount} suggest that the Debye layer is perfectly selective in the limit $(\lambda_D/R)/{\rm Du} \rightarrow 0$.  This can be confirmed as follows:  For a monovalent ionic species, the deviations in counterion (coion) concentration from the centerline potential in the Debye layer are related to $\delta c$ and $n_c$ by $\delta c_{\rm count} = \left( \delta c + |n_c| \right)/2$ ($\delta c_{\rm co} = \left( \delta c - |n_c| \right)/2$).  With these expressions and Eqs. \ref{eqn:electroneutrality} and \ref{eqn:deltac_integral_limit}, we find $\langle \delta c_{\rm count} \rangle = 2|{\rm Du}|c_0$ and $\langle \delta c_{\rm co} \rangle = 0$, confirming that the Debye layer is indeed perfectly selective in this limit.

An upper limit on the selectivity induced by Debye overlap may be obtained by considering the selectivity of the nanopore entrance/exit in the limit $\lambda_D/R \rightarrow \infty$.  In this case, we may impose electrochemical equilibrium across the junction between the interior of the nanopore and the reservoir and local electroneutrality on either side \citep{Bocquet+Charlaix2010}.  The ratio of counterion to total concentration is then found to be \citep{Bocquet+Charlaix2010}
\begin{equation}
\frac{\langle c_{\text{count}} \rangle}{\langle c \rangle} = \frac{\sqrt{\left( \frac{1}{2} \right)^2 + |\text{Du}|^2} + |\text{Du}|}{\sqrt{1 + \left( 2 |\text{Du}| \right)^2}}.
\label{eqn:selectivity_large_lDR}
\end{equation}
This upper limit exceeds the selectivity obtained above for $\lambda_D/R = 2(10^{-1})$ only slightly (Fig. \ref{fig:selectivity}), indicating that the nanopore selectivity rapidly saturates for values of $\lambda_D/R \gtrsim 1/5$.

A lower limit, valid when $\lambda_D/R \ll \text{Du}$, is obtained via Eqns. \ref{eqn:electroneutrality} and \ref{eqn:deltac_integral_limit}.  In this case, we find
\begin{equation}
\frac{\langle c_{\text{count}} \rangle}{\langle c \rangle} = \frac{1}{2} \left( 1 + \frac{2|\text{Du}|}{1 + 2|\text{Du}|} \right) \equiv \frac{1}{2} \left( 1 + \text{STR} \right),
\label{eqn:selectivity_small_lDR}
\end{equation}
where in the second equality we have made use of Eq. \ref{eqn:STR}.  Eq. \ref{eqn:selectivity_small_lDR} shows the deep connection between transport and selectivity and can be made more intuitive as follows.  As shown above, the Debye layer in this limit is perfectly selective, so that the ratio of counterion concentration to total concentration is unity.  As always, the bulk is perfectly non-selective, so that the ratio there is $1/2$.  As the STR is the ratio of surface transport to total transport, the ratio of bulk-to-total transport is $1 -$ STR, and we may estimate the total selectivity on the cross-section based on the partitioning of the ionic transport as
\begin{equation}
\nonumber
\underbrace{\frac{1}{2} \times \left(1 - {\rm STR}\right)}_{\text{bulk}} + \underbrace{1 \times {\rm STR}}_{\text{surface}} = \frac{1}{2} \left( 1 + {\rm STR} \right).
\end{equation}
We thus recover the result given in Eq. \ref{eqn:selectivity_small_lDR}.  This result illustrates that it is the selectivity in the bulk and surface weighted by the dynamic partitioning of the ionic transport that controls local selectivity.

Together, the upper and lower limits on selectivity (Eqs. \ref{eqn:selectivity_large_lDR} and \ref{eqn:selectivity_small_lDR}, respectively), define an envelope of selectivity variation with $\lambda_D/R$ (Fig. \ref{fig:selectivity}).  The conclusion of these results is apparent:  the principal parameter controlling nanopore selectivity is not $\lambda_D/R$ but the local Dukhin number.  This result may be understood as follows:  when $\lambda_D/R$ is small, the local Dukhin number controls both the fraction of the transport in the Debye layer (Eq. \ref{eqn:STR}) and the selectivity of the Debye layer (Eq. \ref{eqn:selectivity_small_lDR}).  A large value of Du means that the majority of the ionic flux is carried within the Debye layer, and that this region is highly selective.  Thus, even though the unselective bulk region takes up the majority of the cross-section, the majority of the transport must pass through the highly selective but relatively small Debye layer.  As this process is controlled by the local Dukhin number and the local partitioning of ionic currents, both of which adjust dynamically, we refer to it as dynamic selectivity.

On the other hand, when $\lambda_D/R$ is large, a significant surface charge (as quantified by the Dukhin number) must still be present to draw counterions into (and exclude coions from) the nanopore and thus render the nanopore highly selective.  This is indicated in the Donnan result for the selectivity (Eq. \ref{eqn:selectivity_large_lDR}).  All together, the result is the dominance of the local Dukhin number in determining the local nanopore selectivity (Fig. \ref{fig:selectivity}).

These results suggest that a nanopore may exhibit significant selectivity when the pore size is comparable to the Dukhin length.  As noted above, the Dukhin length reaches hundreds of nanometers for typical ionic concentrations ($0.1-1$ mM) and surface charge densities ($10 - 100$ mC m$^{-2}$).  This is in strong contrast to traditional ion-selective membranes, which have typically subnanometric pore sizes \citep{Siria_et_al2017}, and indeed to the typical picture of ion-selectivity as occurring only in the presence of strong Debye overlap ($\lambda_D/R \gg 1$) \citep{Bocquet+Charlaix2010}.  

We explore this idea by examining the ion selectivity under an applied concentration difference and voltage in a uniform nanopore, and under an applied voltage in a conical nanopore.

\subsubsection{Transport and dynamic selectivity}
\begin{figure}[h!]
    \centerline{\includegraphics[scale = 0.38]{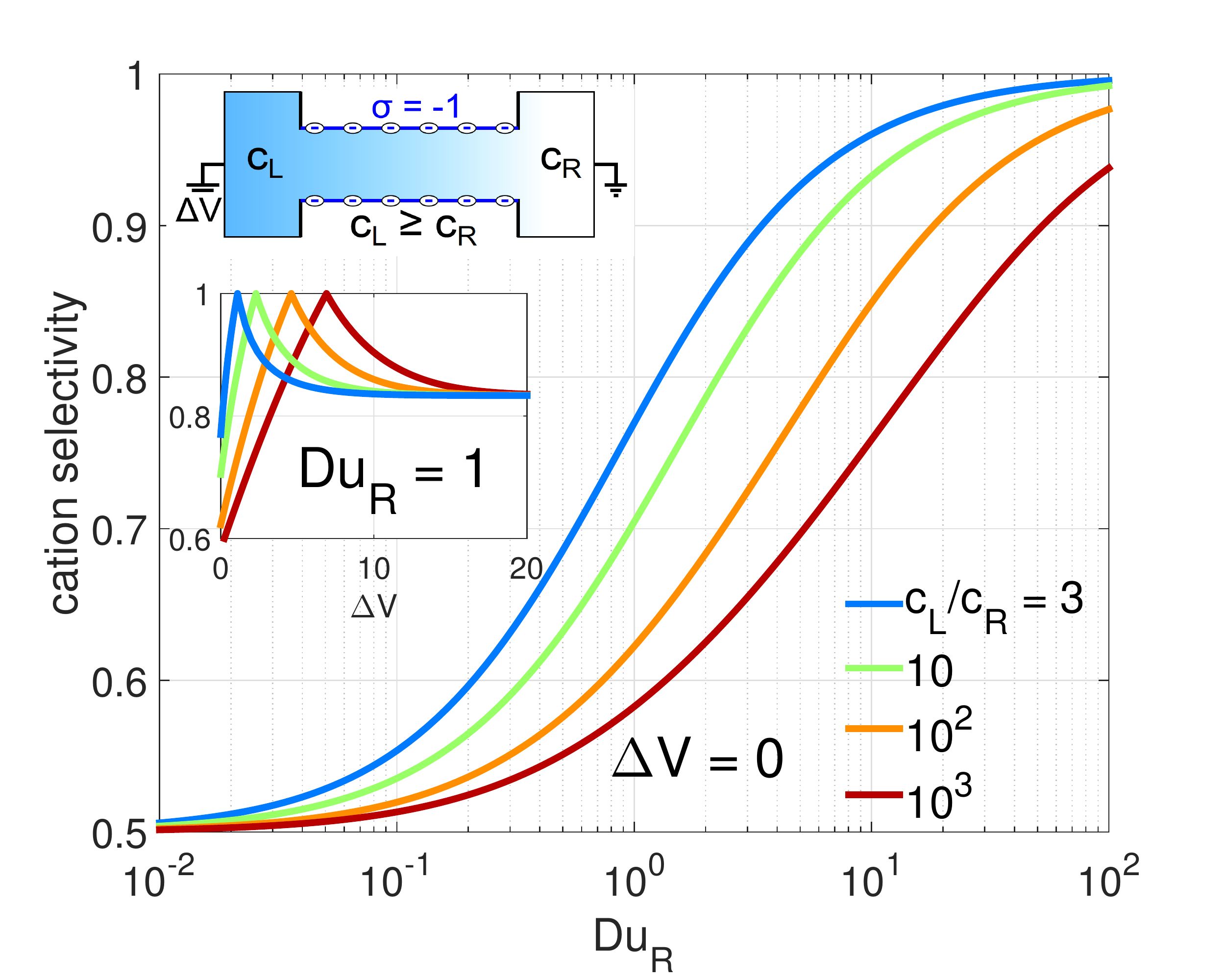}}
    \caption{Cation selectivity for the diffusive flux in the absence of an applied voltage ($\Delta V = 0$) as a function of the larger imposed Dukhin number Du$_R$ corresponding to the smaller reservoir concentration $c_R$.  The curves are colored according to the corresponding value of $c_L/c_R$, as indicated in the legend.  The fluxes are calculated for a nanopore of uniform negative surface charge density and constant radius, as indicated in the schematic in the upper left.  The inset shows the selectivity as a function of voltage applied to the reservoir containing the larger concentration, $c_L$, and for a fixed value of Du$_R = 1$.  The curves are colored according to $c_L/c_R$, as in the main panel, and the voltage is rescaled according to the thermal voltage $k_BT/e \approx 25$ mV.}
    \label{fig:diff_selectivity}
\end{figure}

We first consider a nanopore of uniform negative surface charge density $\sigma$ and constant radius $R$ connecting a left reservoir of concentration $c_L$ and applied voltage $\Delta V$ to a grounded right reservoir of concentration $c_R \leq c_L$ (Fig. \ref{fig:diff_selectivity}, inset). We anticipate here some of the results derived later in order to illustrate the concept of dynamic selectivity.

As detailed below, we may solve the transport equations, Eqs. \ref{eqn:Jsol_final} and \ref{eqn:I_final}, to obtain an implicit algebraic solution for the ionic current and solute flux (Eqs. \ref{eqn:Jsol_soln} and \ref{eqn:IJ_soln}), from which we calculate the ionic flux $J_\pm \equiv (J \pm I)/2$ and the ion selectivities
\begin{equation}
\mathcal{S}_\pm \equiv \frac{|J_\pm|}{|J_+| + |J_-|}.
\label{eqn:ion_flux_selectivity}
\end{equation}
Note from the above definition that the counterion selectivity will fall between $1/2$ and $1$, while the coion selectivity will fall between $0$ and $1/2$, such that $\mathcal{S}_+ + \mathcal{S}_- = 1$.


We show the results of a calculation of the cation (counterion) selectivity in a uniform nanopore under an applied concentration difference in Fig. \ref{fig:diff_selectivity}.  The results are colored according to concentration ratio $c_L/c_R > 1$ and are plotted against the maximum Dukhin number Du$_R \equiv |\sigma|/ec_RR$ imposed at the right end of the nanotube and corresponding to the smaller reservoir concentration $c_R$.  We see that for moderate concentration ratios ($c_L/c_R \leq 10$), a maximum Dukhin number Du$_R$ of order one results in selectivities of $\sim 70 - 80  \%$.

In the inset of Fig. \ref{fig:diff_selectivity} we show the cation selectivity as a function of applied voltage $\Delta V$, again colored according to concentration ratio $c_L/c_R$, and at a fixed Du$_R = 1$.  We recall that the voltage is applied in the high concentration reservoir; this corresponds to the reverse-bias (low conductance) state of the concentration diode.  As we will see below, generically ion selectivity is maximized in a diode when voltage is applied in the reverse-bias direction.  Under an applied voltage, the anion flux is reduced and eventually shut down as the anion chemical potential differential $\Delta \mu_- = {\rm ln}(c_L/c_R) - \Delta V$ decreases and vanishes.  This results in a rapid increase in the cation selectivity as small voltages are applied in the high concentration reservoir and a peak cation selectivity of $100 \%$ when $\Delta V = {\rm ln} (c_L/c_R)$ and $J_- = 0$.  We see in the inset of Fig. \ref{fig:diff_selectivity} that the cation selectivity saturates at a value that is independent of $c_L/c_R$ as $\Delta V$ is increased above ${\rm ln}(c_L/c_R)$ and anions begin to flow from the right to the left reservoir.  This saturation value is given by
\begin{equation}
\mathcal{S}_+(\Delta V \rightarrow +\infty) = \frac{1 + 4{\rm Du}_{\rm max}}{2 + 4{\rm Du}_{\rm max}},
\label{eqn:max_selectivity}
\end{equation}
where Du$_{\rm max}$ is the larger of the two Dukhin numbers imposed on either end of the nanopore; in this case, Du$_{\rm max} =$ Du$_R$.  For the present case, Du$_R = 1$, and the saturation value of the cation selectivity is $\sim 83 \%$.  These results suggest that the nanopore selectivity may be tuned with the application of small external applied voltages.  (Note that $\Delta V$ is rescaled by $k_BT/e \approx 25$ mV such that the maximum plotted voltage in the inset of Fig. \ref{fig:diff_selectivity} $\Delta V = 20$ corresponds to $\sim 500$ mV.)

While the zero-voltage selectivity is not as optimal as traditional ion-selective membranes, which have counterion selectivity ratios $\sim 99\%$, this tradeoff is more than made up for by pore diameters that are one-to-two orders-of-magnitude larger.  

Let us consider two prototypical situations in which such effects could be usefully harnessed. Reverse electrodialysis (RED) is one of a few proposed techniques for the conversion of the osmotic energy associated with the salinity contrast between fresh and saltwater to mechanical energy.  This technique depends on ion-selective diffusive fluxes of the type discussed above and shown in Fig. \ref{fig:diff_selectivity} across stacks of alternating cation- and anion-selective membranes.  The principal limiting factor in commercialization of this process is the low conversion efficiency engendered by the high membrane resistance due to the subnanometric pores in typical ion-selective membranes \citep{Siria_et_al2017}.   Our results suggest that this problem may be circumvented by using large-pore ($10 - 100$ nm) membranes with pore diameters and surface charges tailored to the operating concentrations such that a maximum Dukhin number of order one is achieved.

\begin{figure}[h!]
    \centerline{\includegraphics[scale = 0.43]{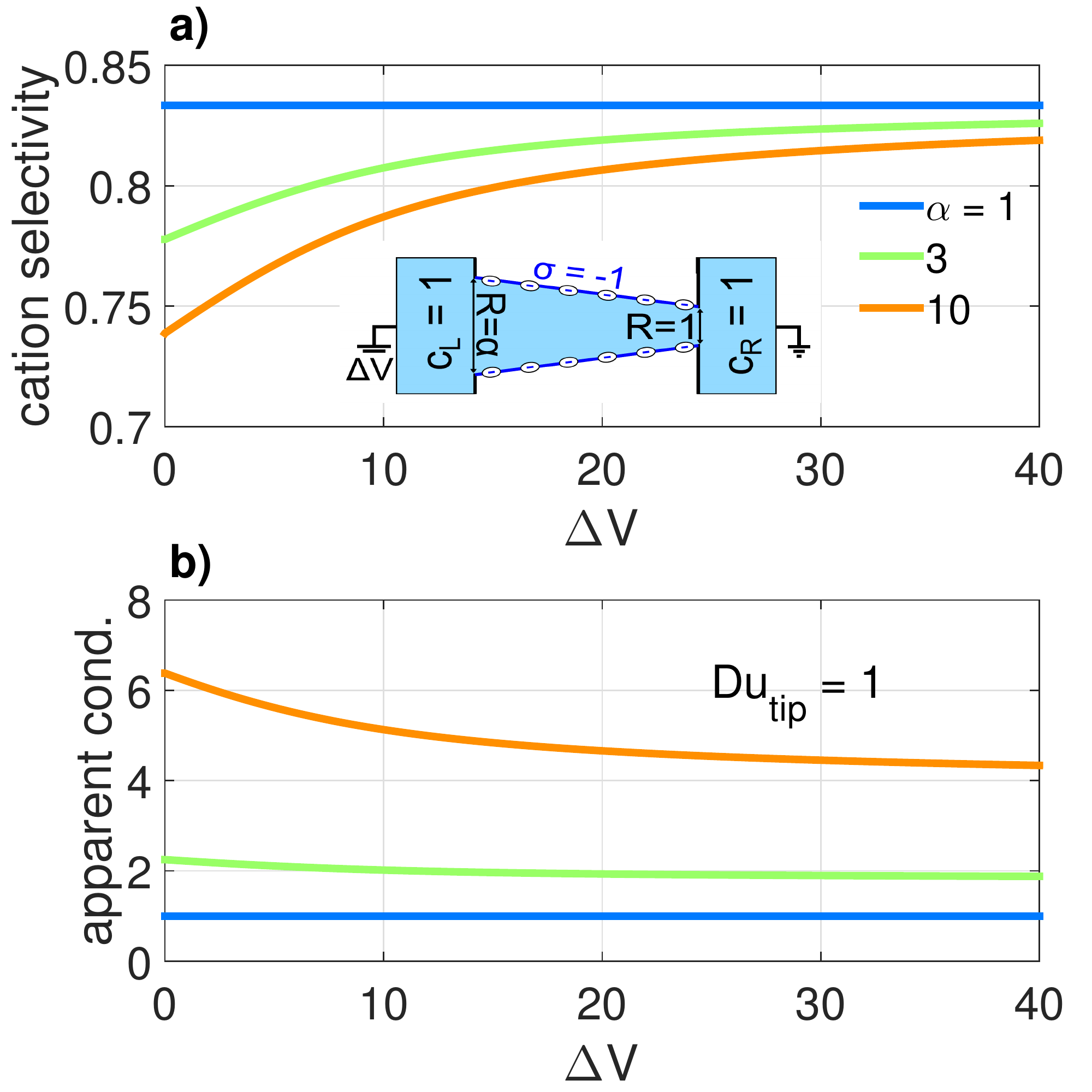}}
    \caption{a) Cation selectivity in a conical nanopore as a function of applied voltage and colored according to the ratio of base and tip radii $\alpha \equiv R_{\rm base}/R_{\rm tip} \geq 1$, as indicated in the legend. b) Apparent conductance $G_{\rm app} \equiv I/\Delta V$ normalized by the conductance of a uniform ($\alpha = 1$) nanopore (Eqs. \ref{eqn:electrophoretic_conductance} through \ref{eqn:G_surf_ep}) as a function of $\Delta V$ and colored according to $\alpha \equiv R_{\rm base}/R_{\rm tip}$, as in a.  The inset in panel a shows a schematic representation of the geometry considered here.  All curves are calculated with a Dukhin number at the tip Du$_{\rm tip} = 1$. The voltage is rescaled by the thermal voltage $k_B T / e \approx 25$ mV, and the variables indicated in the schematic are rescaled according to Table \ref{table:variables}.}
    \label{fig:conical_selectivity}
\end{figure}

Another phenomenon of interest is traditional electrodialysis (ED), in which an electric field is applied across stacks of cation- and anion-selective membranes in order to separate ions from brackish source water.  In this case, what is of interest is the selectivity of the ionic flux induced by an applied voltage in the absence of a concentration differential.  To this end, we first consider as a benchmark the performance (both selectivity and conductance) of a uniform nanopore--i.e., a nanopore with constant (negative) surface charge density and radius; we then compare this to the performance of a conical nanopore having a fixed length, surface charge density, and tip radius $R_{\rm tip}$.  That is, we hold the tip Dukhin number Du$_{\rm tip} \equiv |\sigma|/e c R_{\rm tip}$ fixed while increasing the ratio of base and tip radii $\alpha \equiv R_{\rm base}/R_{\rm tip}$ above unity.  The scenario under consideration is sketched in an inset in Fig. \ref{fig:conical_selectivity}.  Our goal in increasing the opening angle is to improve the conductance compared to the uniform nanopore without a great cost to the nanopore selectivity.

In the case of a uniform nanopore, the transport equations, Eqs. \ref{eqn:Jsol_final} and \ref{eqn:I_final}, are trivially solved for the conductance (given by Eqs. \ref{eqn:electrophoretic_conductance} through \ref{eqn:G_surf_ep_dim}) and the cation selectivity (given by Eq. \ref{eqn:max_selectivity} with Du$_{\rm max} =$ Du$_{\rm tip}$).  Note that in the case of a uniform nanopore the cation selectivity is voltage-independent, as the Dukhin number is equal to Du$_{\rm tip}$ everywhere.  This is indicated by the blue curves in Figs. \ref{fig:conical_selectivity}a, showing the cation selectivity as a function of voltage, and \ref{fig:conical_selectivity}b, showing the apparent conductance $G_{\rm app} \equiv I/\Delta V$ normalized by the uniform nanopore conductance.

The influence of introducing a conical structure (i.e., increasing $\alpha \equiv R_{\rm base}/R_{\rm tip}$ above one) on the cation selectivity is shown in Fig. \ref{fig:conical_selectivity}a.  In the vicinity of $\Delta V = 0$, the selectivity drops rapidly as $\alpha$ is increased, decreasing from the theoretical limit of $\sim 83 \%$ (Eq. \ref{eqn:max_selectivity}) to $\sim 74 \%$ as $\alpha$ is increased from $1$ to $10$; however, as $\Delta V$ is increased, the selectivity again rapidly approaches the theoretical limit for large positive voltages.  We note that the voltage is applied to the larger end of the conical nanopore; this again corresponds to the low conductance and high selectivity configuration of the diode.  We see in Fig. \ref{fig:conical_selectivity}a that a voltage as small as $\sim 1$ V ($\Delta V = 40$) is enough to achieve a selectivity very nearly identical to the uniform nanopore selectivity.

We plot in Fig. \ref{fig:conical_selectivity}b the influence of increasing  $\alpha \equiv R_{\rm base}/R_{\rm tip}$ on the apparent conductance.  We see that the limiting apparent conductance for large positive applied voltages is substantially increased as $\alpha$ is increased.  Increasing $\alpha$ to $3$ is enough to approximately double the conductance, while $\alpha = 10$ results in a conductance that is more than four times larger than the uniform pore conductance.  We will show below that the limiting conductance is related to the uniform nanopore conductance $G_{\rm uni}$ (Eqs. \ref{eqn:electrophoretic_conductance} through \ref{eqn:G_surf_ep_dim}) by
\begin{equation}
G(\Delta V \rightarrow +\infty) = \frac{\alpha - 1}{{\rm ln} \alpha} G_{\rm uni}.
\label{eqn:G_limiting_conical}
\end{equation}

The results for a conical nanopore shown in Fig. \ref{fig:conical_selectivity} and given in Eqs. \ref{eqn:max_selectivity} and \ref{eqn:G_limiting_conical} indicate that 1) substantial selectivity may be achieved in large (i.e., high conductance) uniform radius nanopores if the surface charge and pore radius is tailored to the operating concentrations such that Du$_{\rm tip} \gtrsim 1$ and 2) the conductance may be even further enhanced by introducing a conical shape to the nanopore while holding the tip radius fixed.  Together, these results suggest that, e.g., desalination processes based on ED may be made substantially more efficient by using high surface charge, large, conical nanopores.

\section{Results}

\subsection{The Role of Dynamic Selectivity in the Concentration Diode}
\label{sec:concentration_diode}

In this section, we illustrate the principal role of the Dukhin number in ICR by developing an implicit expression for the IV relationship in a concentration diode using Eqs. \ref{eqn:Jsol_final} and \ref{eqn:I_final}.  For a nanopore of uniform cross-section, these equations are exactly valid cross-sectional integrations of the governing PNP equations in the limits $R/L \rightarrow 0$, $\lambda_D/R \rightarrow 0$, and $(\lambda_D/R)/\text{Du} \rightarrow 0$.  Therefore, the fact that they produce rectified IV curves is strong evidence that $\lambda_D/R \gtrsim \mathcal{O}(1)$ somewhere in the nanopore is not a necessary condition for the existence of ICR.  As the rectification is fundamentally a consequence of the mechanism of dynamic selectivity outlined above, there is no fundamental mechanistic difference between ionic diodes induced by asymmetric geometry, differences in reservoir concentration, or (continuous) asymmetric surface charge profiles induced by, e.g., differences in reservoir pH. In each case, the asymmetry induces an asymmetry in the Dukhin number $|{\rm Du}| = |\sigma|/ecR$ across the nanopore, resulting in an asymmetry in the selectivity of the Debye layer and in the partitioning of the ionic transport between surface and bulk.  We focus here on the concentration diode because we are able to derive an illustrative algebraic solution.

\subsubsection{Fluxes across a concentration diode}
We consider here the same uniform nanopore configuration as described above and shown schematically in the inset of Fig. \ref{fig:diff_selectivity}.  The local Dukhin number along the length of the nanopore is given by
\begin{equation}
\text{Du}(x) = -\frac{\text{Du}_{\text{ref}}}{c},
\label{eqn:Du_diode}
\end{equation}
where the reference Dukhin number $\text{Du}_{\text{ref}} \equiv |\sigma|/e\overline{c}R$ is defined in terms of the magnitude of the uniform surface charge density, the average of the two reservoir concentrations, and the uniform nanotube radius.  We have omitted the subscript zero on the concentration, and we will continue to omit the subscript in what follows, recalling that the indicated total ionic concentrations and electrostatic potentials are centerline values.

We note that, because the surface charge density and nanotube radius do not vary along the length of the nanopore, the variation in the local Dukhin number is determined entirely by the variation in the local concentration.  In addition to the reference Dukhin number, we introduce the local Dukhin numbers on the left and right of the nanotube, $\text{Du}_L \equiv \text{Du}_{\text{ref}}/c_L$ and ${\rm Du}_R \equiv {\rm Du}_{\rm ref}/c_R$, respectively.  We note that the ratio of Dukhin numbers ${\rm Du}_R/{\rm Du}_L \equiv c_L/c_R \geq 1$ is simply the ratio of reservoir concentrations.  With these definitions, we can express the reference Dukhin number as
\begin{equation}
\text{Du}_{\text{ref}} = \frac{2}{\text{Du}_L^{-1} + \text{Du}_R^{-1}}.
\label{eqn:Du_ref_diode}
\end{equation}
We will formulate our results below in terms of the maximum Dukhin number in the system, ${\rm Du}_R$, corresponding to the minimum reservoir concentration, and the concentration ratio ${\rm Du}_R/{\rm Du}_L \equiv c_L/c_R$. Note that, while we do not impose particular values for the reservoir concentrations in our rescaled, dimensionless model, the concentration ratios considered here are consistent with the range of concentrations considered experimentally, typically between 0.1 mM and 1 M.

With the local Dukhin number given in Eq. \ref{eqn:Du_diode}, Eqs. \ref{eqn:Jsol_final} and \ref{eqn:I_final} become
\begin{equation}
\frac{J}{\pi} = - \left[ \frac{dc}{dx} + 2 \text{Du}_{\text{ref}} \left( \frac{d \text{ln} c}{dx} + \frac{d\phi}{dx} \right) \right], \quad \text{and}
\label{eqn:Jsol_diode}
\end{equation}
\begin{equation}
\frac{I}{\pi} = -  \left[ c \frac{d\phi}{dx} + 2 \text{Du}_{\text{ref}} \left( \frac{d \text{ln} c}{dx} + \frac{d\phi}{dx} \right) \right].
\label{eqn:I_diode}
\end{equation}

Solving Eqs. \ref{eqn:Jsol_diode} and \ref{eqn:I_diode} is a straightforward procedure, but before integrating we must take care to ensure that we are imposing appropriate boundary conditions at the nanotube ends.  This is a nontrivial question because the rapid variation in local Dukhin number (from a nonzero value in the nanotube interior to zero in the reservoir) that occurs on either end of the nanotube means that the slowly varying approximation we have used to impose a local PB equilibrium breaks down \citep{Picallo_et_al2013}.  In one-dimensional PNP-based models of nanopore ionic transport, this is typically taken into account by imposing continuity of the electrochemical potential $\mu_{\pm} = \text{ln} c_{\pm} \pm \phi$ across the junction and local electroneutrality on either side \citep{Cervera_et_al2006, Constantin+Siwy2007, Vlassiouk_et_al_acsnano2008, Bocquet+Charlaix2010, Picallo_et_al2013}.  This is justified as follows:  The rapid variation in local geometry and consequently in the local Dukhin number results in localized deviations from equilibrium and electroneutrality and rapid variations in electrostatic potential and ionic concentrations.  The scale of this adjustment region is given by the Debye length.  (See the paper of Shockley \citep{Shockley1949} for an extensive discussion of this point in the equivalent context of semiconductor physics.)  Thus, in the limit $\lambda_D/L \rightarrow 0$, the adjustment region may be treated as a point discontinuity in the ionic concentrations and electrostatic potential.  However, as the ionic flux densities $j_{\pm} = -c_{\pm} \partial_x \mu_{\pm}$ are proportional to the gradient of the electrochemical potential, the discontinuities in ionic concentration and electrostatic potential must be such that continuity of electrochemical potential is maintained, ensuring finite ionic fluxes.  Outside of this adjustment region, the ions again locally equilibrate, and thus local electroneutrality is imposed on either side of the junction.

As we have already assumed a radially uniform electrochemical potential in locally applying a PB equilibrium, imposing continuity of the electrochemical potential between the uniform reservoir values and the values over the entire cross-section in the nanotube interiors reduces to imposing electrochemical continuity between the reservoir and nanopore centerline.  Furthermore, as the ionic charge vanishes at the nanopore centerline, imposition of electroneutrality there amounts to imposing $c_+^{\rm int} = c_-^{\rm int} = c_{\rm int}/2$, where 'int' indicates the value on the interior of the nanopore.  Electrochemical continuity across the junction thus requires ${\rm ln}(c_{\rm res}/2) \pm \phi_{\rm res} = {\rm ln}(c_{\rm int}/2) \pm \phi_{\rm int}$, where 'res' indicates the value in the reservoir.  This condition is satisfied by imposing continuity of the ionic concentration and potential, $c_{\rm res} = c_{\rm int}$ and $\phi_{\rm res} = \phi_{\rm int}$, respectively \citep{Fair+Osterle1971, Rankin+Huang2016}.

We note that 1) this still corresponds to finite discontinuities in the ionic concentrations and electrostatic potential within the Debye layer, and 2) there would be a discontinuity in centerline concentrations and electrostatic potential in the case of Debye overlap, as the ionic charge density would no longer vanish at the nanopore centerline.  In the limit of complete Debye overlap, this corresponds to the formation of a local Donnan equilibrium between the ends of the nanotube and the adjacent reservoirs \citep{Constantin+Siwy2007, Vlassiouk_et_al_acsnano2008, Bocquet+Charlaix2010, Picallo_et_al2013}.

With the above boundary conditions, we can directly integrate Eq. \ref{eqn:Jsol_diode} for the solute flux:
\begin{equation}
\begin{split}
\left( \frac{{\rm Du}_R}{{\rm Du}_L} + 1 \right) \frac{J}{2\pi} = & \left( \frac{{\rm Du}_R}{{\rm Du}_L} - 1 \right) \\
& + 2 {\rm Du}_R \left[ \text{ln} \left( \frac{\text{Du}_R}{\text{Du}_L} \right) + \Delta V \right],
\label{eqn:Jsol_soln}
\end{split}
\end{equation}
where we have used Eq. \ref{eqn:Du_ref_diode} and the definitions of $\text{Du}_L$ and $\text{Du}_R$ to rewrite the result in terms of the maximum Dukhin number Du$_R$ and the ratio Du$_R$/Du$_L \equiv c_L/c_R$.

\begin{figure}[h!]
    \centerline{\includegraphics[scale = 0.38]{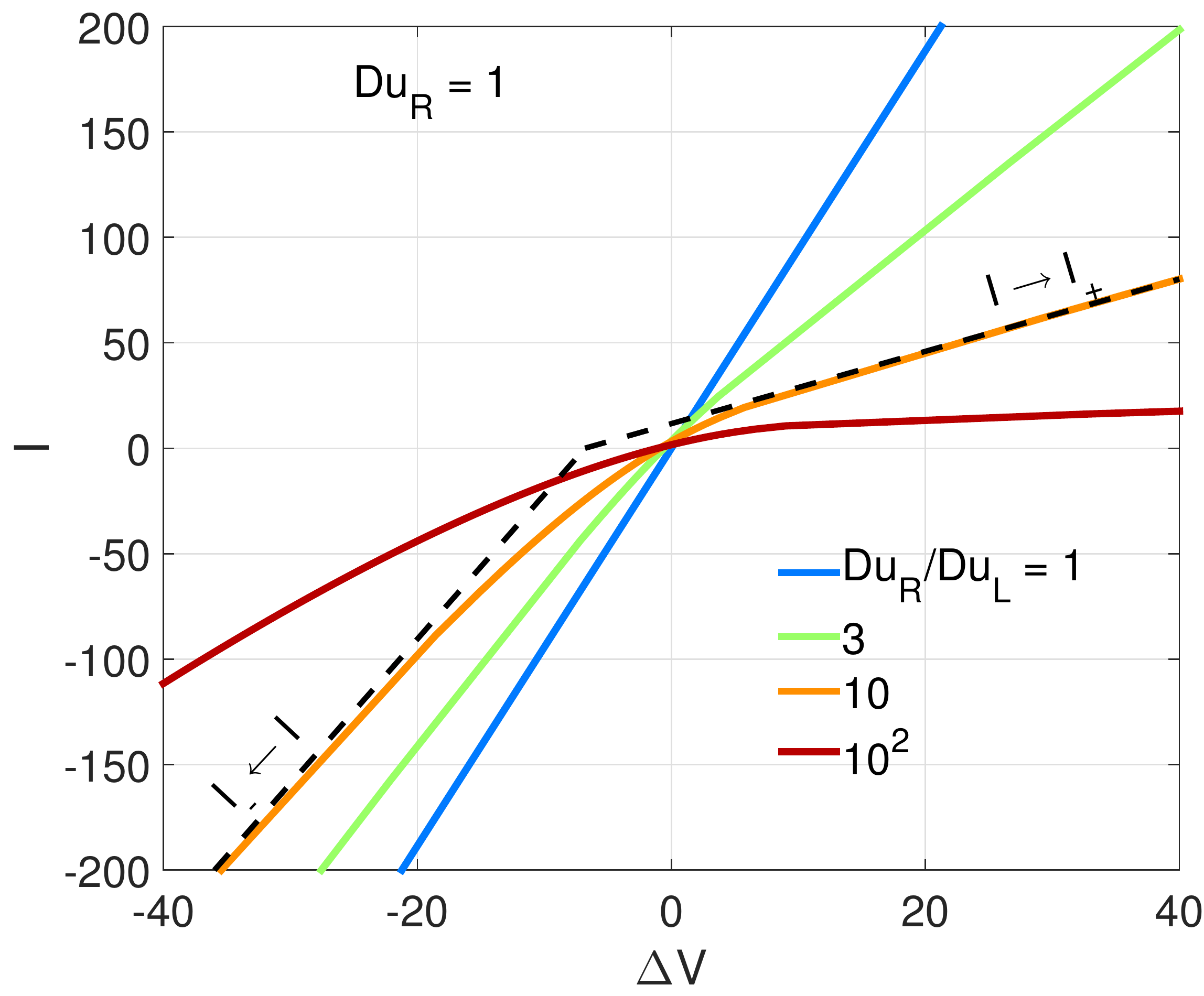}}
    \caption{IV curves obtained from Eqs. \ref{eqn:Jsol_soln} and \ref{eqn:IJ_soln} for a fixed value of $\text{Du}_R = 1$.  The curves are colored according to $\text{Du}_R/\text{Du}_L \equiv c_L/c_R$ as indicated in the legend.  The dashed black lines indicate $I_+$ and $I_-$, the currents obtained in the limit $\Delta V \rightarrow + \infty$ and $-\infty$, respectively, for Du$_R/$Du$_L = 10$.  (See Eqs. \ref{eqn:Iplus} and \ref{eqn:Iminus}.) The voltage is rescaled by the thermal voltage $k_B T / e \approx 25$ mV, and the current is rescaled according to Table \ref{table:variables}.}
    \label{fig:IV}
\end{figure}

Using Eq. \ref{eqn:Jsol_diode}, we solve for $c d\phi/dx$ in terms of $J$ and $dc/dx$, insert the result into Eq. \ref{eqn:I_diode}, and integrate.  The result can be combined with Eq. \ref{eqn:Jsol_soln} to obtain
\begin{equation}
\begin{split}
\text{ln} \left( \frac{\text{Du}_R}{\text{Du}_L} \right) + \Delta V = & \left( 1 + \frac{I}{J} \right) \\
& \times \text{ln} \left[ \frac{\frac{{\rm Du}_R}{{\rm Du}_L} + 2 {\rm Du}_R \left( 1 - \frac{I}{J} \right)}{1 + 2 {\rm Du}_R \left( 1 - \frac{I}{J} \right)} \right].
\label{eqn:IJ_soln}
\end{split}
\end{equation}
Eq. \ref{eqn:IJ_soln} may be used to determine the applied voltage as a function of the ratio $I/J$ for given values of $\text{Du}_R$ and $\text{Du}_R/{\rm Du}_L$, and the result can be combined with Eq. \ref{eqn:Jsol_soln} to determine the solute flux and ionic current as a function of applied voltage.  IV curves obtained using Eqs. \ref{eqn:Jsol_soln} and \ref{eqn:IJ_soln} are plotted in Fig. \ref{fig:IV} for a fixed value of $\text{Du}_R = 1$ and several values of $\text{Du}_R/\text{Du}_L \equiv c_L/c_R \geq 1$.

\subsubsection{Limiting Conductances}
\label{subsec:Gplus_Gminus}

As anticipated, as $\text{Du}_R/\text{Du}_L \rightarrow 1$, representing equal reservoir concentrations, the IV curve becomes progressively more linear, and the conductance approaches a limiting value representing the sum of the bulk and surface electrophoretic contributions \citep{Bocquet+Charlaix2010, Siria_et_al2013}:
\begin{equation}
G = G_{\rm bulk} + G_{\rm surf}, \quad \rm with
\label{eqn:electrophoretic_conductance}
\end{equation}
\begin{equation}
G_{\rm bulk} \equiv \pi, \quad \rm and
\label{eqn:G_bulk_ep}
\end{equation}
\begin{equation}
G_{\rm surf} \equiv 2 \pi {\rm Du}_{\rm ref}.
\label{eqn:G_surf_ep}
\end{equation}
This result is obtained by solving Eqs. \ref{eqn:Jsol_diode} and \ref{eqn:I_diode} with $dc/dx = 0$.  In dimensioned terms, these conductances are given by \citep{Siria_et_al2013}
\begin{equation}
G_{\rm bulk} = \frac{\pi R^2}{L} \frac{e^2 D}{k_B T} c_{\text{res}}, \quad \rm and
\label{eqn:G_bulk_ep_dim}
\end{equation}
\begin{equation}
G_{\rm surf} = \frac{2 \pi R}{L} \frac{e D}{k_B T} |\sigma|,
\label{eqn:G_surf_ep_dim}
\end{equation}
where $c_{\text{res}}$ is the concentration in both reservoirs.  This limiting conductance is indicated by the blue curve in Fig. \ref{fig:IV}.

As a concentration difference is applied ($\text{Du}_R/\text{Du}_L > 1$) and increased, the IV curves become progressively more rectified.  This is due to the asymmetry in selectivity between the left and right end of the nanotube.  The Dukhin number at the right end is held fixed at one, resulting in substantial selectivity for positive coions at that end (Fig. \ref{fig:selectivity}).  On the other hand, as the Dukhin number on the left end is decreased via an increasing reservoir concentration $c_L$, the counterion selectivity at this end rapidly decreases, approaching the non-selective limit for values of $\text{Du}_R/\text{Du}_L > 10$ (Fig. \ref{fig:selectivity}).


In order to fully understand under what conditions ICR occurs, we examine limiting voltage-independent conductances obtained under various conditions.  We have already noted (Eqs. \ref{eqn:electrophoretic_conductance} through \ref{eqn:G_surf_ep_dim}) the limiting conductance when the reservoir concentrations are equal ($\text{Du}_R/\text{Du}_L = 1$).  We next examine the limiting currents and differential conductances $G \equiv \partial I / \partial \Delta V$ when Du$_R$ and Du$_R$/Du$_L$ are held fixed and $\Delta V \rightarrow \pm \infty$, denoted $I_{\pm}$ and $G_{\pm}$, respectively.  From Eq. \ref{eqn:IJ_soln}, we see that the logarithm on the right-hand side must diverge as the voltage diverges for a fixed concentration ratio Du$_R$/Du$_L$.  In the limit $|\Delta V| \rightarrow \infty$, the ionic current and solute flux will be in the same direction, such that the coefficient of the logarithm $1 + I/J > 0$.  Thus, when $\Delta V \rightarrow +\infty$, the argument of the logarithm must diverge.  This requires that $I/J \rightarrow 1 + (2 \text{Du}_R)^{-1} \equiv \text{STR}_R^{-1}$, where STR$_R$ is the surface transport ratio at the right end of the nanopore (Eq. \ref{eqn:STR}).  Combined with Eq. \ref{eqn:Jsol_soln}, this gives for the current and conductance when $\Delta V \rightarrow +\infty$
\begin{equation}
\begin{split}
\left( \frac{2\pi \text{Du}_{\text{ref}}}{\text{STR}_R} \right)^{-1} I_+ = & \frac{1}{2 {\rm Du}_R} \left( \frac{{\rm Du}_R}{{\rm Du}_L} - 1 \right) \\
& + \text{ln} \left( \frac{\text{Du}_R}{\text{Du}_L} \right) + \Delta V, \quad \text{and}
\label{eqn:Iplus}
\end{split}
\end{equation}
\begin{equation}
G_+ = \frac{G_{\rm surf}}{\text{STR}_R} \equiv 2 \pi {\rm Du}_{\rm ref} \frac{1 + 2 {\rm Du}_R}{2 {\rm Du}_R}.
\label{eqn:Gplus}
\end{equation}
Likewise, as $\Delta V \rightarrow -\infty$, Eq. \ref{eqn:IJ_soln} indicates that the argument of the logarithm must vanish, and thus $I/J \rightarrow \text{STR}_L^{-1}$.  By the same procedure we find
\begin{equation}
\begin{split}
\left( \frac{2\pi \text{Du}_{\text{ref}}}{\text{STR}_L} \right)^{-1} I_- = & \frac{1}{2 {\rm Du}_R} \left( \frac{{\rm Du}_R}{{\rm Du}_L} - 1 \right) \\
& + \text{ln} \left( \frac{\text{Du}_R}{\text{Du}_L} \right) + \Delta V, \quad \text{and}
\label{eqn:Iminus}
\end{split}
\end{equation}
\begin{equation}
G_- = \frac{G_{\rm surf}}{\text{STR}_L} \equiv 2 \pi {\rm Du}_{\rm ref} \frac{1 + 2 {\rm Du}_L}{2 {\rm Du}_L}.
\label{eqn:Gminus}
\end{equation}
These limiting currents are indicated in Fig. \ref{fig:IV} for $\text{Du}_R = 1$ and $\text{Du}_R/\text{Du}_L = 10$.

As $\text{Du}_R$ becomes smaller for a fixed value of the ratio $\text{Du}_R/\text{Du}_L$, the voltage magnitude that leads to significant concentration accumulation or depletion becomes larger.  This is because the asymmetry in nanopore selectivity between either end of the nanopore becomes weaker (Fig. \ref{fig:selectivity}).  This means that, for an experimentally feasible range of applied voltages, the IV curve linearizes as $\text{Du}_R$ is decreased.  From Eq. \ref{eqn:IJ_soln}, we see that as $\text{Du}_R \rightarrow 0$ for a fixed value of $\text{Du}_R/\text{Du}_L$, $I/J \rightarrow \Delta V / \text{ln} (\text{Du}_R/\text{Du}_L)$.  This is true so long as the numerator and denominator of the logarithm are $\gg 0$, which, from our discussion above, requires that the voltage magnitude not be too large.  We find from this limit and Eq. \ref{eqn:Jsol_soln} a limiting conductance
\begin{equation}
G \rightarrow \frac{ 2 \left( \frac{\text{Du}_R}{\text{Du}_L} - 1 \right) }{\left( \frac{\text{Du}_R}{\text{Du}_L} + 1 \right) \text{ln} \left( \frac{\text{Du}_R}{\text{Du}_L} \right)} G_{\rm bulk},
\label{eqn:G_DuR_small}
\end{equation}
valid for fixed $\text{Du}_R/\text{Du}_L$ when $\text{Du}_R \ll 1$.  We note that the prefactor in Eq. \ref{eqn:G_DuR_small} approaches unity and $G \rightarrow G_{\rm bulk}$ in the limit $\text{Du}_R/\text{Du}_L \rightarrow 1$, as it must.

Finally, we examine the limit $\text{Du}_R \rightarrow \infty$ for fixed $\text{Du}_R/\text{Du}_L$.  In this case, the entire nanopore becomes perfectly selective for counterions (Fig. \ref{fig:selectivity}), and the IV curve again linearizes.  Unlike in the case that $\text{Du}_R \ll 1$, however, we find that the IV curve is linear irrespective of the magnitude of the applied voltage for large $\text{Du}_R$.  From Eq. \ref{eqn:STR}, we note that, as $\text{Du} \rightarrow \infty$, $\text{STR} \rightarrow 1$.  Thus, as $\text{Du}_R \rightarrow \infty$ for fixed $\text{Du}_R/\text{Du}_L$, $\text{STR}_L \rightarrow \text{STR}_R \rightarrow 1$, and, from Eqns. \ref{eqn:Gplus} and \ref{eqn:Gminus}, we find $G_+ \rightarrow G_- \rightarrow G_{\rm surf}$.  In this case, the conductance is dominated by the (concentration-independent) surface contribution.

Each of the limiting conductances discussed above, and the conditions under which they obtain, are listed in Table \ref{table:conductances}.

\begin{table}
 \begin{center}
  \begin{tabular}{cc}
	 condition					& G \\ \hline
	\addlinespace[1mm]
       $c_{\rm max}/c_{\rm min} = 1$		& $G_{\rm bulk} + G_{\rm surf}$ \\
	\addlinespace[1mm]
       $\text{Du}_{\rm max} \ll 1$			& $\frac{ 2 ( c_{\rm max}/c_{\rm min} - 1) }{( c_{\rm max}/c_{\rm min} + 1 ) \text{ln} \left( c_{\rm max}/c_{\rm min} \right)} \times G_{\rm bulk}$ \\
	\addlinespace[1mm]
       $\text{Du}_{\rm max} \rightarrow \infty$	& $G_{\rm surf}$ \\
	\addlinespace[1mm]
       $|\Delta V| \rightarrow \infty$		& \begin{tabular}{c}$G_{\rm min} = \frac{1 + 2 {\rm Du}_{\rm max}}{2 {\rm Du}_{\rm max}} \times G_{\rm surf}$ \\ \addlinespace[1mm] $G_{\rm max} = \frac{1 + 2 {\rm Du}_{\rm min}}{2 {\rm Du}_{\rm min}} \times G_{\rm surf}$ \end{tabular} \\
  \end{tabular}
  \caption{Limiting conductances for the concentration diode and the conditions under which they obtain.  The results are written generically in terms of the maximum and minimum reservoir concentrations, $c_{\rm max}$ and $c_{\rm min}$, respectively, and the corresponding maximum and minimum Dukhin numbers, Du$_{\rm max} \equiv |\sigma|/ec_{\rm min}R$ and Du$_{\rm min} \equiv |\sigma|/ec_{\rm max}R$, imposed on either end of the nanopore.  $G_{\rm max}$ and $G_{\rm min}$ are, respectively, the maximum and minimum conductances obtained as $|\Delta V| \rightarrow \infty$.  (See Eqs. \ref{eqn:Gplus}, \ref{eqn:Gminus} and \ref{eqn:Gmax}, \ref{eqn:Gmin}.)  The bulk ($G_{\rm bulk}$) and surface ($G_{\rm surf}$) electrophoretic condutances are given in Eqs. \ref{eqn:G_bulk_ep} and \ref{eqn:G_surf_ep}, respectively.}
  \label{table:conductances}
 \end{center}
\end{table}

\subsubsection{Rectification Ratio}
\label{subsec:rect_ratio}

\begin{figure}[h!]
    \centerline{\includegraphics[scale = 0.38]{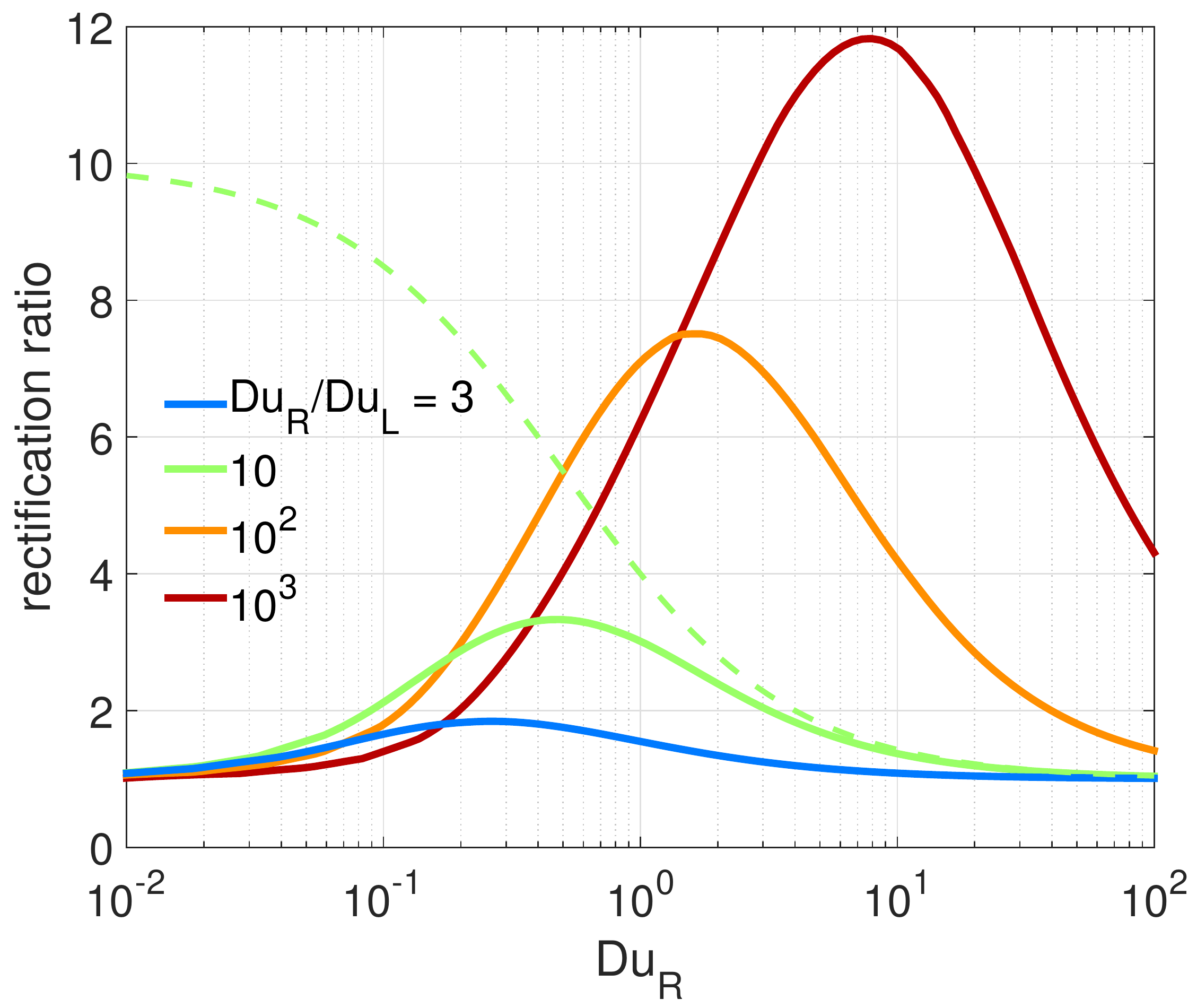}}
    \caption{Rectification ratio obtained from Eqs. \ref{eqn:Jsol_soln} and \ref{eqn:IJ_soln} and evaluated at $|\Delta V| = 40$ ($\approx 1 \, \text{V}$), as a function of $\text{Du}_R$.  The curves are colored according to $\text{Du}_R/\text{Du}_L = c_L/c_R$ as indicated in the legend.  The dashed line indicates the theoretical maximum rectification ratio, STR$_R$/STR$_L$, valid in the limit $|\Delta V| \rightarrow \infty$, for Du$_R$/Du$_L = 10$.}
    \label{fig:rect_ratio}
\end{figure}

We are now in a position to discuss the rectification ratio, defined as
\begin{equation}
\text{rectification ratio} \equiv \frac{|I(-|\Delta V|) - I(\Delta V = 0)|}{|I(+|\Delta V|) - I(\Delta V = 0)|}.
\label{eqn:rect_ratio}
\end{equation}
We plot this ratio in Fig. \ref{fig:rect_ratio} as a function of Du$_R$ for fixed concentration ratios Du$_R$/Du$_L = c_L/c_R$. In general, the rectification ratio is a function of the voltage magnitude $|\Delta V|$; for definiteness, we take $|\Delta V| = 40$, which corresponds to a dimensioned applied voltage of approximately $1 \, \text{V}$.

The rectification ratios display a peak for a finite value of $\text{Du}_R$ (Fig. \ref{fig:rect_ratio}).  This peak grows and is shifted to higher values of Du$_R$ as Du$_R$/Du$_L$ is increased.  The value of Du$_R$ corresponding to the peak rectification ratio Du$_R^{\rm peak}$ is of order one over much of the parameter space ($0.2 < {\rm Du}_R^{\rm peak} < 2$ for $3 \leq {\rm Du}_R/{\rm Du}_L \leq 10^2$).

The fact that the rectification ratio shows a peak at a finite value of Du$_R$ can be understood as follows:  By taking the ratio of the limiting conductances valid when $|\Delta V| \rightarrow \infty$ (Eqs. \ref{eqn:Gplus} and \ref{eqn:Gminus}), we see that there is an upper limit on the maximum rectification, given by $G_-/G_+ = {\rm STR}_R/{\rm STR}_L$ and valid in the limit $|\Delta V| \rightarrow \infty$.  As Du$_R \rightarrow 0$ for a fixed value of Du$_R$/Du$_L$, this ratio reaches a maximum value equal to the concentration ratio Du$_R$/Du$_L$.  However, in this case, the local Dukhin number is much smaller than one everywhere and the nanopore is therefore only weakly selective for counterions (Fig. \ref{fig:selectivity}).  It thus takes very large voltages to engender significant ion accumulation or depletion, voltages much larger than the practical upper limit in nanofluidic experiments ($\sim 1$ V), and the IV curve is effectively linearized.

On the other hand, when Du$_R \rightarrow \infty$ (with Du$_R$/Du$_L$ fixed), STR$_R \rightarrow$ STR$_L \rightarrow 1$, indicating that the IV curve is strictly linear in this limit, irrespective of the voltage magnitude.  This is because the local Dukhin number is everywhere much larger than one and the entirety of the ionic transport is carried within the Debye layer.  There is therefore no repartitioning of the transport between bulk and surface and no gradient in ion selectivity along the length of the nanopore.  In this case, the conductance is given by the (concentration-independent) surface conductance (Eq. \ref{eqn:G_surf_ep_dim}).  Thus, the location of Du$_R^{\rm max}$ represents a compromise between the non-selective (Du$_R \rightarrow 0$) and perfectly selective (Du$_R \rightarrow \infty$) limits.

The occurrence of a maximum rectification ratio for a finite value of Du$_R$ and a fixed value of the ratio Du$_R$/Du$_L$ is exactly analogous to the common observation of a maximum rectification ratio for a finite concentration or surface charge density and a fixed geometry in conical diodes, \citep[e.g.][]{Ai_et_al2010, Cervera_et_al2006, Vlassiouk_et_al2009, Zhou_et_al2011}.  In that case, the ratio of Dukhin numbers is fixed by the ratio of base and tip radii, while the variation of concentration or surface charge results in a variation of the maximum Dukhin number occuring at the tip of the conical nanopore.  As in the concentration diode, the location of the maximum is determined by a compromise between the non-selective (high concentration or low surface charge) and perfectly selective (low concentration or high surface charge) limits.  We will discuss the role of dynamic selectivity in diodes induced by asymmetric geometry (and surface charge density distributions) below.

\subsection{Dynamic Selectivity and Limiting Conductances in Generic Diodes}
\label{sec:generic_diodes}

\subsubsection{Rectification in Geometric and Concentration Diodes}
\label{subsec:rect_general}

\begin{figure}[h!]
    \centerline{\includegraphics[scale = 0.37]{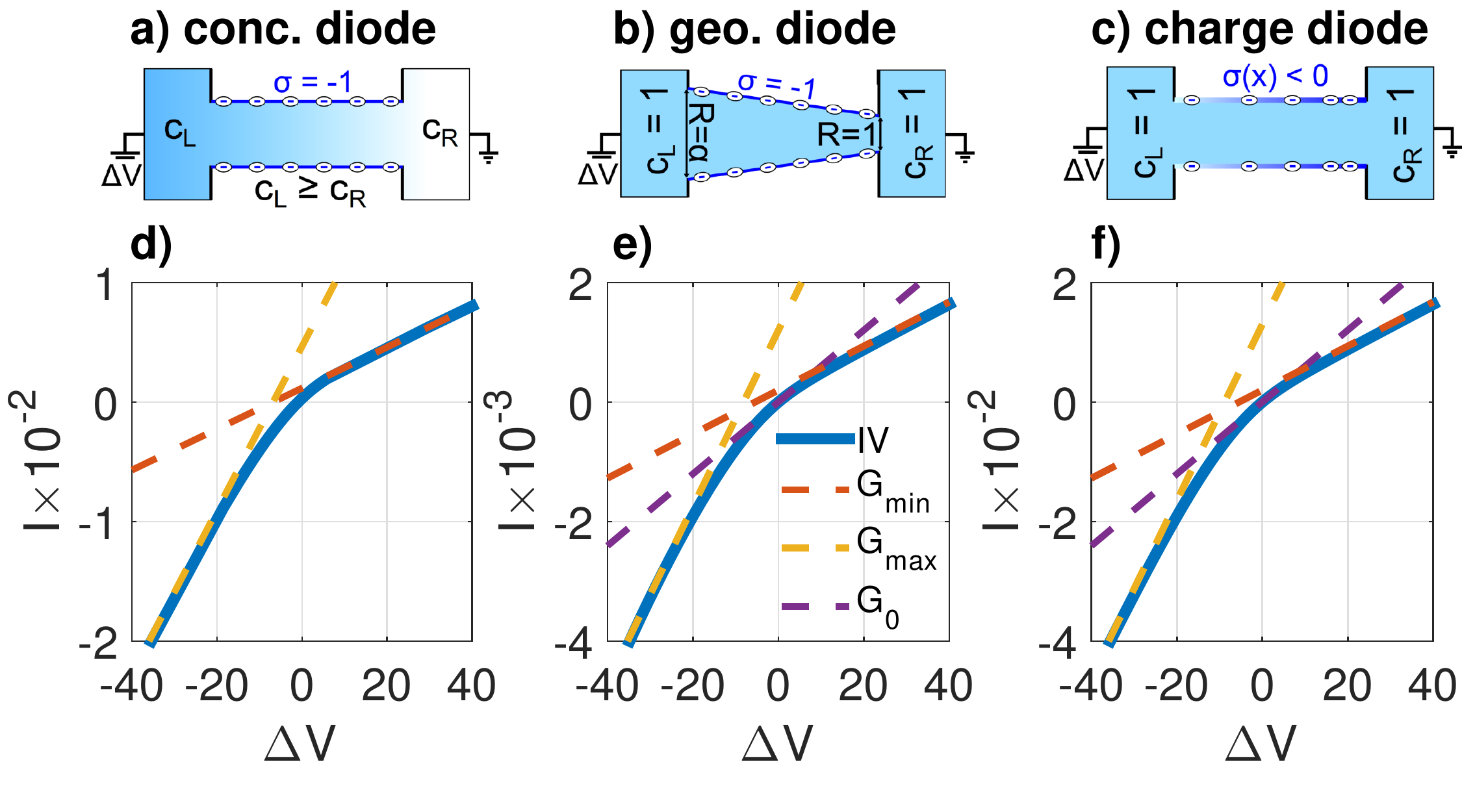}}
    \caption{a-c)  Schematics of diodes induced by a) unequal reservoir concentrations, b) asymmetric geometry, and c) asymmetric surface charge distribution.  d-f)  IV curves obtained for Du$_R = 1$ and Du$_L = 0.1$ for d) the concentration diode shown in a, e) the geometric diode shown in b, and f) the charge diode shown in c.  In panels d-f, the dashed yellow and red lines show the maximum and minimum conductances obtained when $|\Delta V| \rightarrow \infty$ and calculated according to Eqs. \ref{eqn:Gmax} and \ref{eqn:Gmin}, respectively.  In panels e and f, the dashed purple line indicates the linear response conductance valid in the vicinity of $\Delta V = 0$ for the charge and geometric diodes and calculated according to Eq. \ref{eqn:G0_general}. The voltage is rescaled by the thermal voltage $k_B T / e \approx 25$ mV, and the current as well as the variables indicated in the schematics are rescaled according to Table \ref{table:variables}.}
    \label{fig:IV_compare}
\end{figure}

From our understanding of the role of the Dukhin number in controlling local selectivity, and of the mechanism of dynamic selectivity in controlling rectification, we conclude that ICR is generically a consequence of inequality of the Dukhin numbers imposed on either end of a nanopore, irrespective of whether that asymmetry is induced by a difference in reservoir concentrations, an asymmetric geometry, or an asymmetric surface charge density distribution (or any combination thereof).  This is corroborated by the results shown in Fig. \ref{fig:IV_compare}, where we compare an IV curve obtained from the above solution for a concentration diode (Eqs. \ref{eqn:Jsol_soln} and \ref{eqn:IJ_soln}, Figs. \ref{fig:IV_compare}a and d) to numerical solutions of the transport equations (Eqs. \ref{eqn:Jsol_final} and \ref{eqn:I_final}) for the IV curve in a geometric (Figs. \ref{fig:IV_compare}b and e) and a charge diode (Figs. \ref{fig:IV_compare}c and f).  In order to illustrate rectification induced by an asymmetric geometry, we assume a linear variation in the nanopore radius from the maximum radius (minimum Dukhin number) on the left to the minimum radius (maximum Dukhin number) on the right.  The surface charge density is taken to be fixed and negative, and the reservoir concentrations are taken to be equal.  This configuration is shown schematically in Fig. \ref{fig:IV_compare}b.

Likewise, we illustrate rectification induced by a continuous, asymmetric surface charge profile by imposing a negative surface charge density whose magnitude varies linearly from a minimum density (minimum Dukhin number) on the left to a maximum density (maximum Dukhin number) on the right.  In this case, we impose a constant nanopore radius and equal reservoir concentrations.  This configuration is shown schematically in Fig. \ref{fig:IV_compare}c.  In all three cases, we take Du$_L = 0.1$ and Du$_R = 1$.  We immediately see from Figs. \ref{fig:IV_compare}d through f that the qualitative structure of the rectified IV curve is essentially the same over a given range of voltage ($-40 \leq \Delta V \leq +40$ here).  Indeed, in each case the rectification ratio (Eq. \ref{eqn:rect_ratio}) is $\sim 3$ for $|\Delta V| = 40$.  The qualitative similarity of the IV curves obtained in these three different configurations illustrates the equivalence of the mechanism resulting in rectification.

\subsubsection{General Expressions for the Limiting Conductances and Selectivities When $|\Delta V| \rightarrow \infty$}
\label{subsec:limitingG}

In this section, we show that the above expressions obtained for the limiting conductances when $|\Delta V| \rightarrow \infty$ (Eqs. \ref{eqn:Gplus} and \ref{eqn:Gminus}) obtained for the concentration diode are particular examples of general expressions relating the minimum (maximum) Dukhin number imposed at one end of the nanopore to the maximum (minimum) conductance obtained for large imposed voltages.  We also derive expressions for the limiting ion-selectivities when $|\Delta V| \rightarrow \infty$.  In the case of the geometric and charge diodes, these correspond to the minimum and maximum selectivities obtainable by varying the applied voltage.  These expressions are valid for concentration, geometric, and charge diodes (or any combination thereof), as illustrated in Fig. \ref{fig:IV_compare} for the limiting conductances.  In the course of this discussion, we illustrate general features of the evolution of the Dukhin number profile in the nanopore interior as a function of applied voltage, further illustrating the principal role of dynamic selectivity in the accumulation or depletion of concentration in the nanopore interior and hence in ICR.

The general expressions for limiting conductances in generic diodes, along with those derived below for the linear response near $\Delta V = 0$ in geometric and charge diodes, will allow observations of rectified IV curves to be related to, for example, the surface charge density.  The surface charge density is difficult to estimate directly and is often estimated by observing the saturation of the conductance at the surface-dominated value for low concentrations, \citep[e.g.][]{Stein_et_al2004, Karnik_et_al2005, Secchi_et_al2016_prl}.  However, the inference of surface charge from conductance measurements typically relies on a linear response, in which case the relation between surface charge and the saturating conductance at low concentration is known analytically, \citep[e.g.][]{Stein_et_al2004, Bocquet+Charlaix2010, Siria_et_al2013}.  It is not clear \textit{a priori} how this framework may be extended to, e.g., conical nanopores, where ICR is inherent to the IV response below a certain concentration.  To our knowledge, general analytical results for the relationship between surface charge and conductance do not exist in the literature for rectified IV curves, except in certain specialized scenarios, \citep[e.g.][]{Picallo_et_al2013}.

Starting from Eqs. \ref{eqn:Jsol_final} and \ref{eqn:I_final}, we may write the transport equations generally as
\begin{equation}
\frac{J}{\pi R^2} = - \left[ \frac{dc}{dx} + 2 |{\rm Du}| \left( \frac{dc}{dx} - {\rm S} c \frac{d\phi}{dx} \right) \right], \quad {\rm and}
\label{eqn:Jsol_general}
\end{equation}
\begin{equation}
\frac{I}{\pi R^2} = - \left[ c\frac{d\phi}{dx} - {\rm S} 2 |{\rm Du}| \left( \frac{dc}{dx} - {\rm S} c \frac{d \phi}{dx} \right) \right].
\label{eqn:I_general}
\end{equation}
Note that we have assumed nothing about the sign of the surface charge S (except that it does not change along the length of the nanopore) or the nature of the variation of the local Dukhin number $|{\rm Du}|$.

As $|\Delta V| \rightarrow \infty$, the solute flux will be dominated by the surface electrophoretic mass transport, $J/\pi R^2 \sim {\rm S} 2 |{\rm Du}| c d\phi/dx = {\rm S} 2 {\rm Du}_{\rm ref} (|\sigma|/R) d\phi/dx$.  Integrating in $x$, we find
\begin{equation}
J = -{\rm S} \left( \int_L \frac{{\rm dx}}{2 \pi R |\sigma|} \right)^{-1} {\rm Du}_{\rm ref} \Delta V.
\label{eqn:Jsol_highV}
\end{equation}

In order to obtain a condition on the flux ratio $I/J$ that holds in the limit $|\Delta V| \rightarrow \infty$, we solve Eqs. \ref{eqn:Jsol_general} and \ref{eqn:I_general} for the concentration gradient $dc/dx$ in terms of $I/J$, the local Dukhin number, and the (divergent) solute flux:
\begin{equation}
\frac{dc}{dx} = -{\rm S} \frac{ \frac{I}{J} + {\rm S} \left( 1 + \frac{1}{2|{\rm Du}|} \right) }{2 + \frac{1}{2|{\rm Du}|} } \frac{J}{\pi R^2}.
\label{eqn:J_dcdx}
\end{equation}
On physical grounds, the concentration gradient cannot diverge everywhere in the nanopore interior, even in the limit that $|\Delta V| \rightarrow \infty$.  Accordingly, the prefactor in Eq. \ref{eqn:J_dcdx} must vanish as the solute flux diverges.  This requires that $I/J \rightarrow -{\rm S}(1 + 1/2|{\rm Du}|) \equiv -{\rm S}/{\rm STR}$, where in the second equality we have made use of Eq. \ref{eqn:STR}.   As the ratio $I/J$ is spatially uniform at steady state, this condition requires the Dukhin number in the nanopore interior to approach a uniform value, which we will designate ${\rm Du}_u$.  Likewise, we designate the corresponding surface transport ratio STR$_u \equiv 2{\rm Du}_u/(1 + 2{\rm Du}_u )$.  With this result for the flux ratio $I/J$ and Eq. \ref{eqn:Jsol_highV} we find for the current
\begin{equation}
I = \left( \int_L \frac{{\rm dx}}{2 \pi R |\sigma|} \right)^{-1} \frac{{\rm Du}_{\rm ref}}{{\rm STR}_u} \Delta V.
\label{eqn:I_highV}
\end{equation}

The mechanism of concentration accumulation/depletion is driven by the gradient in Dukhin number induced by the asymmetry between the maximum (Du$_{\rm max}$) and minimum (Du$_{\rm min}$) Dukhin numbers imposed on either end of the nanopore.  Thus, for very strong applied voltages, the accumulation (depletion) will cease when the concentration everywhere is such that the uniform Dukhin number in the interior is equal to Du$_u$ = Du$_{\rm min}$ (Du$_{\rm max}$).  At one end of the nanopore, the concentration gradient then must diverge to match the divergence in the solute flux (Eq. \ref{eqn:Jsol_highV}) while allowing the Dukhin number to deviate from its uniform interior value and adjust to the appropriate boundary condition.

\begin{figure}[h!]
    \centerline{\includegraphics[scale = 0.37]{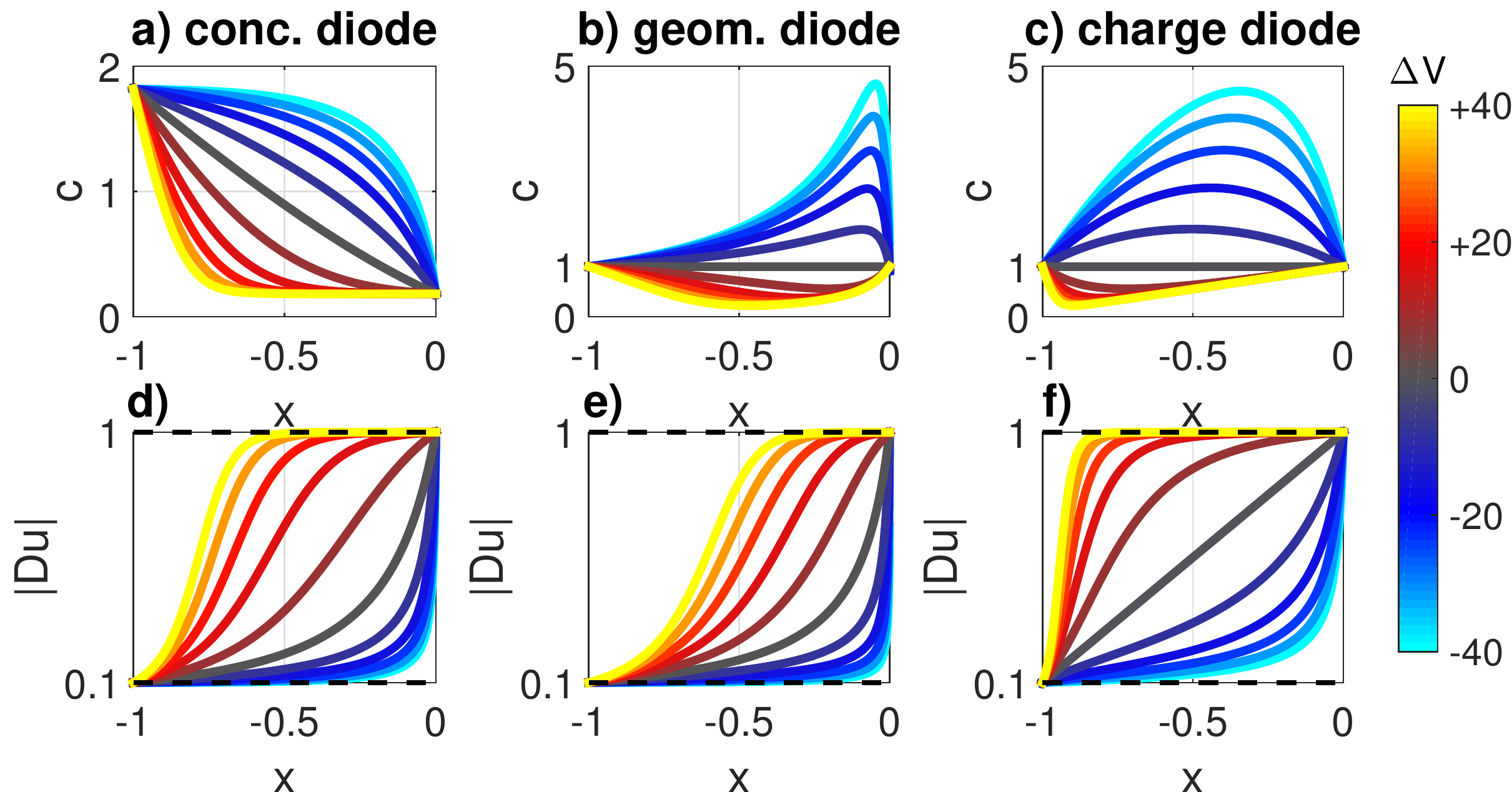}}
    \caption{a-c) Profiles of total ionic concentration along the length of the a) concentration diode shown schematically in Fig. \ref{fig:IV_compare}a, b) the geometric diode (Fig. \ref{fig:IV_compare}b), and c) the charge diode (Fig. \ref{fig:IV_compare}c).  d-f) The corresponding profiles of local Dukhin number for the concentration (d), geometric (e), and charge (f) diodes. The dashed black lines in d-f indicate the minimum and maximum imposed Dukhin numbers at either end of the nanopore, Du$_L = 0.1$ and Du$_R = 1$, respectively. In all panels, the curves are colored according to the applied voltage, as indicated in the colorbar on the right. The voltages are rescaled by the thermal voltage $k_B T / e \approx 25$ mV, and the concentrations are rescaled by the mean of the reservoir concentrations.}
    \label{fig:conc_profiles}
\end{figure}

This mechanism is illustrated in Fig. \ref{fig:conc_profiles}:  In Figs. \ref{fig:conc_profiles}a through c, we show the profiles of centerline concentration as a function of the applied voltage for the concentration (Fig. \ref{fig:conc_profiles}a), geometric (Fig. \ref{fig:conc_profiles}b), and charge (Fig. \ref{fig:conc_profiles}c) diodes shown schematically in Figs. \ref{fig:IV_compare}a through c, respectively.  In the case of the concentration diode, Eq. \ref{eqn:J_dcdx} may be integrated to obtain an implicit expression for the concentration profile, while for the geometric and charge diodes it is necessary to solve the transport equations (Eqs. \ref{eqn:Jsol_final} and \ref{eqn:I_final}) numerically to obtain the concentration profiles.  As for the IV curves shown in Figs. \ref{fig:IV_compare}d through f, the profiles are calculated for Du$_L = 0.1$ and Du$_R = 1$.  The qualitative structure of the concentration profiles is quite different in the three configurations considered; however, in each case, there is increasing depletion (accumulation) of concentration in the nanopore interior for increasing magnitude positive (negative) voltage.  Note that the sign of the voltage resulting in accumulation/depletion would be inverted for a positive surface charge, rather than the negative surface charge considered here.

However, in Figs. \ref{fig:conc_profiles}d through f, we see that the evolution of the local Dukhin number profiles $|{\rm Du}(x)|$ with applied voltage are strikingly similar in the three configurations, even though the concentration profiles are quite different.  In each case, an increasing magnitude positive (negative) voltage results in a growing region in the nanopore interior where $|{\rm Du}| \approx {\rm Du}_{\rm max} = {\rm Du}_R$ ($|{\rm Du}| \approx {\rm Du}_{\rm min} = {\rm Du}_L$).  Fig. \ref{fig:conc_profiles} illustrates the key role of the local Dukhin number in controlling the accumulation/depletion of concentration in the nanopore and hence in ICR.  It also illustrates that, in the extreme limit that $|\Delta V| \rightarrow \infty$, the Dukhin number will approach a uniform value equal to the maximum or minimum Dukhin number imposed at one end of the nanopore.  (Whether it approaches Du$_{\rm max}$ or Du$_{\rm min}$ depends on the sign of the applied voltage and the surface charge density.)  Thus, denoting the maximum and minimum limiting conductances as $G_{\rm max}$ and $G_{\rm min}$, respectively, we find
\begin{equation}
G_{\rm max} = \frac{G_{\rm surf}}{{\rm STR}_{\rm min}} \equiv G_{\rm surf} \frac{1 + 2 {\rm Du}_{\rm min}}{2 {\rm Du}_{\rm min}}, \quad \rm and
\label{eqn:Gmax}
\end{equation}
\begin{equation}
\begin{split}
G_{\rm min} = \frac{G_{\rm surf}}{{\rm STR}_{\rm max}} \equiv G_{\rm surf} \frac{1 + 2 {\rm Du}_{\rm max}}{2 {\rm Du}_{\rm max}}.
\label{eqn:Gmin}
\end{split}
\end{equation}
In these equations, $G_{\rm surf}$ is the surface conductance obtained when Du$_{\rm ref} \rightarrow \infty$ and only the surface terms in Eq. \ref{eqn:I_general} are relevant \citep{Frament+Dwyer2012}:
\begin{equation}
\begin{split}
G_{\rm surf} &=  \left( \int_L \frac{\rm dx}{2 \pi R |\sigma|} \right)^{-1} {\rm Du}_{\rm ref} \\
&\xrightarrow{\text{dim.}} \frac{eD}{k_B T} \left( \int_L \frac{\rm dx}{2 \pi R |\sigma|} \right)^{-1},
\end{split}
\label{eqn:Gsurf_general}
\end{equation}
where we have redimensionalized in the second line.  The limiting conductances predicted using Eqs. \ref{eqn:Gmax} and \ref{eqn:Gmin} are shown in Figs. \ref{fig:IV_compare}d through f (dashed yellow and red lines).

We note that, by setting $|\sigma| \equiv R \equiv 1$ and identifying ${\rm Du}_{\rm min} = {\rm Du}_L$ and ${\rm Du}_{\rm max} = {\rm Du}_R$, we recover Eqs. \ref{eqn:Gplus} and \ref{eqn:Gminus} from Eqs. \ref{eqn:Gmin} and \ref{eqn:Gmax}, respectively.

Figs. \ref{fig:conc_profiles}d through f show that the Dukhin number has not fully approached a uniform value everywhere in the nanopore interior even for $|\Delta V| = +40$; however, Figs. \ref{fig:IV_compare}d through f indicate that the differential conductance is roughly equal to its limiting value for $|\Delta V| \gtrsim 10 - 20$ ($\sim 250-500$ mV).  We note, however, that the voltage necessary to reach the limiting conductance depends on both the maximum Dukhin number in the system Du$_{\rm max}$ and the asymmetry in Dukhin numbers, quantified by the ratio Du$_{\rm max}$/Du$_{\rm min}$.  (See Fig. \ref{fig:rect_ratio} and related discussion.)  Thus, we note that care must be taken in applying Eqs. \ref{eqn:Gmax} and \ref{eqn:Gmin} to experimental IV curves.  This difficulty can be avoided by instead fitting the surface charge to the linear response conductance $G_0$ obtained in the vicinity of $\Delta V = 0$; we will obtain an analytical expression for $G_0$ in the following section.

Using the asymptotic expressions for the solute flux (Eq. \ref{eqn:Jsol_highV}) and ionic current (Eqs. \ref{eqn:I_highV} through \ref{eqn:Gsurf_general}) we can calculate the co-/counterion fluxes $J_{\rm co/count} = (J \pm {\rm S}I)/2$ and derive expressions for the counterion selectivities $\mathcal{S}_{\rm count} \equiv |J_{\rm count}|/(|J_{\rm co}| + |J_{\rm count}|)$ in the limits $\Delta V \rightarrow \pm \infty$.  The results are
\begin{equation}
\mathcal{S}_{\rm count}^{\rm max} = \frac{1 + 4 {\rm Du}_{\rm max}}{2 + 4 {\rm Du}_{\rm max}} \quad \rm (reverse-bias);
\label{eqn:max_count_selectivity}
\end{equation}
\begin{equation}
\mathcal{S}_{\rm count}^{\rm min} = \frac{1 + 4 {\rm Du}_{\rm min}}{2 + 4 {\rm Du}_{\rm min}} \quad \rm (forward-bias).
\label{eqn:min_count_selectivity}
\end{equation}
Eq. \ref{eqn:max_count_selectivity} confirms Eq. \ref{eqn:max_selectivity}.  Furthermore, we see that the ion-selectivity is maximized in the reverse-bias (low conductance) configuration, as previously noted.  The coion selectivity is of course given by the relation $\mathcal{S}_{\rm co} + \mathcal{S}_{\rm count} = 1$.

Finally, before leaving this section, we derive Eq. \ref{eqn:G_limiting_conical} relating the conical conductance in the reverse-bias configuration to the conductance of a uniform nanopore having the same surface charge density and a constant radius equal to the tip radius.  Eq. \ref{eqn:Gmin} holds in the reverse-bias configuration.  We perform the integration in Eq. \ref{eqn:Gsurf_general} with $|\sigma| = 1$ and imposing a linear variation in the radius to find $G_{\rm surf} = 2 \pi {\rm Du}_{\rm tip} (\alpha - 1)/{{\rm ln} \alpha}$.  We combine this result with Eq. \ref{eqn:Gmin}, recognizing that $2 \pi {\rm Du}_{\rm tip} \times (1 + 2 {\rm Du}_{\rm tip})/2 {\rm Du}_{\rm tip} = \pi + 2 \pi {\rm Du}_{\rm tip} \equiv G_{\rm uni}$, the uniform reference nanopore conductance (Eqs. \ref{eqn:electrophoretic_conductance} through \ref{eqn:G_surf_ep}), to find
\begin{equation}
G_{\rm min} = \frac{\alpha - 1}{{\rm ln} \alpha} G_{\rm uni}.
\label{eqn:G_limiting_conical2}
\end{equation}
This confirms Eq. \ref{eqn:G_limiting_conical}.

\subsubsection{Conductance in the Vicinity of $\Delta V = 0$}
\label{subsec:linear_response}

In the case of the concentration diode, the imposed concentration difference means that there is a difference in electrochemical potentials between the reservoirs for at least one of the ionic species irrespective of the applied voltage.  However, for the geometric and charge diodes equilibrium obtains when $\Delta V = 0$, and we can linearize about this equilibrium to obtain an expression for the differential conductance $G_0 \equiv \partial I / \partial \Delta V |_{\Delta V = 0}$ in the vicinity of $\Delta V = 0$.  The equilibrium state is characterized by $\Delta V = 0 \Longrightarrow J_{\rm co} = J_{\rm count} = 0$ and the concentration, electrostatic potential, and electrochemical potential profiles $c \equiv 1$, $\phi \equiv 0$, and $\mu_{\rm co} \equiv \mu_{\rm count} \equiv - {\rm ln}2$, respectively.  We introduce a perturbative forcing $\delta V \ll 1$, which induces fluxes $\delta J_{\rm co}$ and $\delta J_{\rm count}$.  The applied voltage perturbs the concentration and electrostatic potential profiles such that $c \rightarrow 1 + c'$ and $\phi \rightarrow 0 + \phi'$, with $c' = 0$ on either end of the nanopore and $\phi'$ varying between $\delta V$ on the left and $0$ on the right end of the nanopore.  The electrochemical potentials become $\mu_{\rm co} \rightarrow - {\rm ln}2 + c' + {\rm S} \phi'$ and $\mu_{\rm count} \rightarrow - {\rm ln}2 + c' - {\rm S} \phi'$, from which we define $\mu_{{\rm co}/{\rm count}}' \equiv c' \pm {\rm S} \phi'$.

We linearize Eqs. \ref{eqn:Jco} and \ref{eqn:Jcount} to find
\begin{equation}
\frac{\delta J_{\rm co}}{\pi R^2} = -\frac{1}{2} \frac{d \mu_{\rm co}'}{dx}, \quad \rm and
\label{eqn:Jco_lin_diff}
\end{equation}
\begin{equation}
\frac{\delta J_{\rm count}}{\pi R^2} = -\left( \frac{1}{2} + 2 {\rm Du}_{\rm ref} \frac{|\sigma|}{R} \right) \frac{d \mu_{\rm count}'}{dx}.
\label{eqn:Jcount_lin_diff}
\end{equation}
Integration of Eqs. \ref{eqn:Jco_lin_diff} and \ref{eqn:Jcount_lin_diff} along the length of the nanopore gives
\begin{equation}
\delta J_{\rm co} = +{\rm S} \frac{1}{2} \left( \int_L \frac{\rm dx}{\pi R^2} \right)^{-1} \delta V, \quad \rm and
\label{eqn:Jco_lin}
\end{equation}
\begin{equation}
\delta J_{\rm count} = - {\rm S} \frac{1}{2} \left[ \int_L \frac{\rm dx}{\pi R^2 \left( 1 + 4 {\rm Du}_{\rm ref} |\sigma|/R \right)} \right]^{-1} \delta V.
\label{eqn:Jcount_lin}
\end{equation}
From these results we may calculate the conductance at $\Delta V = 0$ as $G_0 \equiv \delta I / \delta V \equiv {\rm S} (\delta J_{\rm co} - \delta J_{\rm count})/\delta V$.  We find
\begin{equation}
\begin{split}
G_0 &= \overbrace{\left( 2 \int_L \frac{\rm dx}{\pi R^2} \right)^{-1}}^{G_0^{\rm bulk}} \\
 &+ \underbrace{\left[ 2 \int_L \frac{\rm dx}{\pi R^2 (1 + 4 {\rm Du}_{\rm ref} |\sigma|/R)} \right]^{-1}}_{G_0^{\rm surf}},
\label{eqn:G0_general}
\end{split}
\end{equation}
where we have partitioned the result into bulk and surface contributions.  The conductances predicted from Eq. \ref{eqn:G0_general} are shown in Fig. \ref{fig:IV_compare} for a diode induced by a linear variation in i) nanopore radius (Fig. \ref{fig:IV_compare}e) and ii) surface charge density (Fig. \ref{fig:IV_compare}f).

For the sake of illustration, we derive an explicit expression for $G_0$ in a conical nanopore with a linearly varying radius and uniform surface charge density.  Many studies have looked at ICR in such nanopores, \citep[e.g.][]{Ai_et_al2010, Cervera_et_al2006, Constantin+Siwy2007, Kovarik_et_al2009, Lan_et_al2011, Laohakunakorn+Keyser2015, Liu_et_al2007, Vlassiouk+Siwy2007, White+Bund2008, Woermann2003, Siwy+Fulinski2002}, and our result will be useful in relating rectified IV curves to surface charge densities in conical nanopores.

We take the radius to vary linearly between a maximum at the base of the conical nanopore, $R_{\rm base}$, and a minimum at the tip, $R_{\rm tip}$.  This gives for the magnitude of the radial slope $|dR/dx| = \alpha - 1$ (in rescaled variables), where $\alpha \equiv R_{\rm base}/R_{\rm tip}$, as introduced above.  We insert this into Eq. \ref{eqn:G0_general}, along with the condition that the surface charge density is uniform $|\sigma| = 1$, and evaluate the integrals to find
\begin{equation}
G_0^{\rm bulk} = \pi \frac{\alpha}{2}, \quad \rm and
\label{eqn:G0_conical_bulk}
\end{equation}
\begin{equation}
G_0^{\rm surf} = \frac{ \alpha - 1}{{\rm ln} \left( \frac{1 + 4{\rm Du}_{\rm tip}}{1 + 4{\rm Du}_{\rm tip}/\alpha} \right)} 2 \pi {\rm Du}_{\rm tip}.
\label{eqn:G0_conical_surf}
\end{equation}
In the above, we have recognized that Du$_{\rm ref} =$ Du$_{\rm tip} \equiv |\sigma|/ec_{\rm res}R_{\rm tip}$, the Dukhin number defined in terms of the uniform surface charge density magnitude, reservoir concentration, and tip radius.  Redimensioning Eqs. \ref{eqn:G0_conical_bulk} and \ref{eqn:G0_conical_surf}, we find
\begin{equation}
G_0^{\rm bulk} = \frac{\pi R_{\rm base} R_{\rm tip}}{2L} \frac{e^2 D}{k_B T} c_{\rm res}, \quad \rm and
\label{eqn:G0_conical_bulk_dim}
\end{equation}
\begin{equation}
G_0^{\rm surf} = \frac{ \alpha - 1}{{\rm ln} \left( \frac{1 + 4{\rm Du}_{\rm tip}}{1 + 4{\rm Du}_{\rm tip}/\alpha} \right)} \frac{2 \pi R_{\rm tip}}{L} \frac{e D}{k_B T} |\sigma|.
\label{eqn:G0_conical_surf_dim}
\end{equation}
Note that, in the limit of a uniform nanopore $\alpha \rightarrow 1$, the sum of Eqs. \ref{eqn:G0_conical_bulk} (\ref{eqn:G0_conical_bulk_dim}) and \ref{eqn:G0_conical_surf} (\ref{eqn:G0_conical_surf_dim}) agrees with the sum of Eqs. \ref{eqn:G_bulk_ep} (\ref{eqn:G_bulk_ep_dim}) and \ref{eqn:G_surf_ep} (\ref{eqn:G_surf_ep_dim}), as it must.

We find that the dependence of the surface conductance on the surface charge (through the terms proportional to Du$_{\rm tip}$ appearing in the logarithm) is more complicated than a linear proportionality, indicating that the results for the conductance in a conical nanopore reported in\citep{Frament+Dwyer2012}, for example, cannot be na\"{i}vely applied to the linear response conductance obtained for $|\Delta V| \ll k_B T / e$ (in dimensioned terms).  The conductance derived in \citep{Frament+Dwyer2012} indicates a linear proportionality between the surface conductance and the magnitude of the surface charge density, and we obtain the expression for the surface conductance given therein from Eq. \ref{eqn:G0_conical_surf_dim} only in the limit that 4Du$_{\rm tip} \gg \alpha$.

\section{Discussion}
\label{sec:discussion}

\subsection{A Reanalysis of ICR Data in the Literature}

Our results suggest that inequality of the Dukhin numbers imposed on either end of a nanopore is the only criterion for the occurrence of ICR.  Further, they suggest that Du $\sim 1$ is a (rough) criterion for the maximization of rectification.  To this end, we have reinterpreted data for conical nanopores in the literature in terms of the (maximal) tip Dukhin number (Table \ref{table:lit_data}).  We see that substantial rectification may be obtained even when the minimum radius is two-to-three orders-of-magnitude larger than the Debye length, but that peak rectification consistently corresponds to Du $\sim 1$, consistent with our theoretical description of dynamic selectivity and its role in ICR.

\begin{table}
 \begin{center}
  \begin{tabular}{c|c|c}
	 peak rect. ratio	& $\lambda_D/R_{\rm tip}$	& Du$_{\rm tip}$  \\ \hline
	 $10$\citep{He_et_al2017}			& $6(10^{-4})$ 				& $1.3$ \\
 	$15$\citep{Lin_et_al2018}			& $0.014$ 				& $0.38$ \\
	$4.6$\citep{Cervera_et_al2006}			& $0.046$ 				& $0.41$ \\
	$2.4$\citep{Jubin_et_al2018}			& $0.082$ 				& $1.3$ \\
	$6.5$\citep{Kovarik_et_al2009}			& $0.17$				& $4.8$ \\
	$5.2$\citep{Cervera_et_al2006}			& $0.33$				& $2.9$ \\
	$1.9$\citep{White+Bund2008}			& $0.61$				& $1.0$ \\
  \end{tabular}
  \caption{Maximum rectification ratios in conical nanopores for fixed Dukhin ratios (Du$_{\rm max}$/Du$_{\rm min} = R_{\rm base}/R_{\rm tip}$) along with corresponding values of $\lambda_D/R_{\rm tip}$ and Du$_{\rm tip}$, as estimated from the literature. (The superscripts on the peak values of the rectification ratio indicate the corresponding reference.)  Note that the peak rectification ratios cannot be directly compared as they are not all calculated at the same reference voltage magnitude.  (The reference voltage magnitudes range between $400$ mV and $2$ V.)}
  \label{table:lit_data}
 \end{center}
\end{table}

In selecting experimental and numerical data from the literature, we searched for any rectification data that was obtained by imposing a \textit{continuous} variation in concentration, geometry, and/or surface charge.  Immediately, this excludes data obtained in charge diodes containing a discontinuity in the magnitude and/or sign of the surface charge, \citep[e.g.][]{Li_et_al2013, Nguyen_et_al2010, Karnik_et_al2007}.  The important distinction between diodes containing discontinuities in the local Dukhin number or the sign of the local surface charge and those considered here will be discussed in the following section.  We additionally searched for rectification ratios (either directly reported or inferred from reported IV curves) that displayed a local maximum as the maximum Dukhin number was varied (via variations either in reservoir concentration or surface charge density) while the Dukhin ratio was held fixed.  In the end, all of the data that fit our criteria were found to come from conical nanopores.

\subsection{A Note on the Distinction Between Intrinsic and Extrinsic Diodes}
\label{subsec:intrinsic}

Briefly, we note that we have been concerned here with the ICR induced by continuous variations in the local Dukhin number and in the presence of surface charge of a single sign.  This is in contrast to both classical bipolar diodes, containing regions of both positive and negative surface charge, \citep[e.g.][]{Picallo_et_al2013, Nguyen_et_al2010, Constantin+Siwy2007, Vlassiouk+Siwy2007}, and unipolar diodes, \citep[e.g.][]{Karnik_et_al2007, Wang_et_al2007, Li_et_al2013}, containing regions of zero and nonzero surface charge.  We term the latter intrinsic diodes, as in this case the zone of depletion or accumulation is localized to the intrinsic discontinuity in either the local Dukhin number or the sign of the surface charge.  We term the type of diodes considered here extrinsic diodes, in contrast to the previous terminology and in recognition of the fact that, in this case, the rectification is due to an imposed inequality in the Dukhin numbers on either end of the nanopore, rather than an intrinsic discontinuity.

Intrinsic diodes are typically found to exhibit much stronger rectification \citep{Li_et_al2013, Vlassiouk+Siwy2007}, due to the presence of a localized intrinsic accumulation/depletion zone.  Picallo \textit{et al.} \citep{Picallo_et_al2013} showed analytically that, in the limit of high surface charge (Du $\rightarrow \infty$), bipolar diodes exhibit ideal Shockley behavior, typical of classical p-n junction semiconductor diodes \citep{Shockley1949}.  In this case, the current is described by $I = I_{\rm sat} \left[ 1 - {\rm exp} \left( -e \Delta V / k_B T \right) \right]$, where $I_{\rm sat}$ is the finite saturation current obtained for large positive (reverse-biased) voltages.  This is in strong contrast to the behavior of extrinsic diodes as detailed above, in which two finite limiting conductances are observed for large positive or negative voltages, and it is further notable because it illustrates that rectification in intrinsic diodes is maximized as Du $\rightarrow \infty$, rather than being washed out.

\section{Conclusions and Perspectives}

The principal conclusion of this work is that the key parameter controlling nanopore ion-selectivity, through the mechanism of dynamic selectivity, is the Dukhin number Du $\equiv |\sigma|/ecR$; the ratio of Debye length to nanopore radius $\lambda_D/R$ is of secondary importance.  As the Dukhin length $\ell_{\rm Du} \equiv |\sigma|/ec$ can reach values of hundreds of nanometers for typical surface charge densities and ionic concentrations, this suggests that significant ionic selectivity can be obtained for large ($10 - 100$ nm) nanopores.  This result has allowed us to rationalize the observation that significant ionic current rectification is routinely observed even when the nanopore radius is one-to-three orders of magnitude larger than the Debye length.  In doing so, we have obtained several general formulae for limiting and linear response conductances in ionic diodes, which will allow researchers to relate IV measurements in asymmetric systems to unknown parameters such as the surface charge density.

Crucially, this mechanism suggests the possibility of designing large, conical pore ion-selective membranes.  The tip radius and surface charge density of such membrane nanopores can be tailored to the operating concentrations in order to obtain significant ion-selectivity ($80 - 90 \%$) while achieving orders-of-magnitude larger conductances than those obtained in traditional (subnanometric) ion-selective membranes.  The implications for the design of much more efficient osmotic energy conversion (RED) and desalination/filtration (ED) devices is profound, as the key limiting factor in commercialization of such technologies is the poor efficiency due to low membrane conductance.  Future work will focus on development of prototypical devices to demonstrate high-efficiency energy conversion and desalination in large nanopores.

\begin{acknowledgement}

A.R.P. acknowledges funding from the European Union's Horizon 2020 Framework Program/European Training Program 674979 - NanoTRANS. A.S. acknowledges funding from the European Union's Horizon 2020 Framework Programme/ERC Starting Grant agreement number 637748 - NanoSOFT. L.B. acknowledges funding from ANR project Neptune and from the European Union's Horizon 2020 Framework Programme/ERC Advanced Grant agreement number 785911 - Shadoks.

\end{acknowledgement}

\bibliography{references_nanofluidics}

\end{document}